\documentclass[fleqn,usenatbib]{mnras}

\usepackage{newtxtext,newtxmath,color}
\usepackage{graphics, setspace, bm, graphicx}
\allowdisplaybreaks

\newcommand{\mathsym}[1]{{}}
\newcommand{\unicode}[1]{{}}

\newcommand{\Sg}{\Sigma}
\newcommand{\ag}{\alpha}
\newcommand{\bg}{\beta}
\newcommand{\cg}{\gamma}

\newcommand{\dg}{\delta}
\newcommand{\Dg}{\Delta}
\newcommand{\kg}{\kappa}

\newcommand{\om}{\omega}
\newcommand{\pom}{\varpi}
\newcommand{\lam}{\lambda}
\newcommand{\pd}{\partial}

\newcommand{\der}{{\rm d}}

\newcommand{\tom}{\tilde \omega}
\newcommand{\ta}{\tilde a}
\newcommand{\taout}{\tilde a_{\rm out}}

\newcommand{\cV}{\mathcal{V}}

\newcommand{\ain}{a_{\rm in}}
\newcommand{\aout}{a_{\rm out}}
\newcommand{\ein}{e_{\rm in}}

\newcommand{\meps}{\epsilon}

\newcommand{\Md}{M_{\rm d}}

\newcommand{\Mf}{M_{\rm f}}
\newcommand{\Ef}{E_{\rm f}}
\newcommand{\af}{a_{\rm f}}
\newcommand{\ef}{e_{\rm f}}

\newcommand{\Teff}{T_{\rm eff}}

\newcommand{\cH}{\mathcal{H}}

\newcommand{\Ms}{M_\star}
\newcommand{\Rs}{R_\star}

\newcommand{\Mh}{M_\bullet}
\newcommand{\Rt}{R_{\rm t}}
\newcommand{\dotMfb}{\dot M_{\rm f}}
\newcommand{\tf}{t_{\rm f}}
\newcommand{\tfo}{t_{\rm f0}}
\newcommand{\Eorb}{E_{\rm orb}}

\newcommand{\sg}{\sigma}
\newcommand{\Rg}{R_{\rm g}}
\newcommand{\ard}{a_{\rm rd}}

\newcommand{\beps}{\bar \varepsilon}
\newcommand{\epsInt}{\langle \bar \varepsilon \rangle}

\newcommand{\Msun}{{\rm M}_\odot}
\newcommand{\Rsun}{{\rm R}_\odot}

\newcommand{\bMh}{\bar M_\bullet}
\newcommand{\bMs}{\bar M_\star}
\newcommand{\bRs}{\bar R_\star}

\newcommand{\cO}{\mathcal{O}}

\newcommand{\rp}{r_{\rm p}}

\newcommand{\Lx}{L_{\rm X}}
\newcommand{\Luv}{L_{\rm UV/Opt.}}

\newcommand{\be}{\begin{equation}}
\newcommand{\ee}{\end{equation}}

\usepackage[T1]{fontenc}

\DeclareRobustCommand{\VAN}[3]{#2}
\let\VANthebibliography\thebibliography
\def\thebibliography{\DeclareRobustCommand{\VAN}[3]{##3}\VANthebibliography}

\usepackage{graphicx}	

\title[Eccentric TDE Disks]{Eccentric Tidal Disruption Event Disks around Supermassive Black Holes: Dynamics and Thermal Emission}

\author[Zanazzi \& Ogilvie]{
J.~J. Zanazzi$^{1,2}$\thanks{E-mail: jzanazzi@cita.utoronto.ca (JJZ)}
and Gordon I. Ogilvie$^{2}$
\\
$^{1}$Canadian Institute for Theoretical Astrophysics, University of Toronto, 60 St. George Street, Toronto, Ontario, M5S 1A7, Canada\\
$^{2}$Department of Applied Mathematics and Theoretical Physics, University of Cambridge, Centre for Mathematical Sciences, Wilberforce Road,\\ Cambridge
CB3 0WA, UK
}

\date{Accepted XXX. Received YYY; in original form ZZZ}

\pubyear{2020}

\begin{document}
\label{firstpage}
\pagerange{\pageref{firstpage}--\pageref{lastpage}}
\maketitle

\begin{abstract}
After the Tidal Disruption Event (TDE) of a star around a SuperMassive Black Hole (SMBH), if the stellar debris stream rapidly circularizes and forms a compact disk, the TDE emission is expected to peak in the soft X-ray or far Ultra-Violet (UV).  The fact that many TDE candidates are observed to peak in the near UV and optical has challenged conventional TDE emission models.  By idealizing a disk as a nested sequence of elliptical orbits which communicate adiabatically via pressure forces, and {are} heated by energy dissipated during the circularization of the nearly parabolic debris streams, we investigate the dynamics and thermal emission of highly eccentric TDE disks, including the effect of General-Relativistic apsidal precession from the SMBH.  We calculate the properties of uniformly precessing, apsidally aligned, and highly eccentric TDE disks, and find highly eccentric disk solutions exist for realistic TDE properties (SMBH and stellar mass, periapsis distance, etc.).  Taking into account compressional heating (cooling) near periapsis (apoapsis), we find our idealized eccentric disk model can produce emission consistent with the X-ray and UV/Optical luminosities of many optically bright TDE candidates.  Our work attempts to quantify the thermal emission expected from the shock-heating model for TDE emission, and finds stream-stream collisions are a promising way to power optically bright TDEs.
\end{abstract}




\section{Introduction}

When a bright transient event occurs near the center of an otherwise quiescent galaxy, with a smooth powerlaw decline in luminosity over a timescale of months to years and negligible corresponding color or blackbody temperature evolution, the most popular interpretation is that we are witnessing a Tidal Disruption Event (TDE) of a star from the galaxy's central SuperMassive Black Hole (SMBH; see review by \citealt{Komossa(2015)}).  TDEs occur when the tidal force exerted on the star by the SMBH exceeds the star's self-gravity, causing the stellar material to be launched on highly elliptical ballistic trajectories.  The stellar debris eventually accretes onto the SMBH, powering the observed luminous transients \citep{Rees(1988),EvansKochanek(1989)}.

An outstanding problem in astrophysics is the low effective temperature of many of the observed TDEs.  Since the earliest TDE models predicted a compact accretion disk which formed quickly from circularization of the stellar debris, with highly super-Eddington accretion rates, the first estimates of TDE emission predicted Eddington luminosities ($\sim 10^{44}-10^{45} \, {\rm erg}\,{\rm s}^{-1}$), compact photosphere radii ($\sim 10^{13} \, {\rm cm}$), and high effective temperatures ($\gtrsim {10^5} \, {\rm K}$; \citealt{Cannizzo(1990),Ulmer(1999)}).  Many TDEs detected a decade or two later had these expected properties \citep[e.g.][]{Bade(1996),KomossaBade(1999),Greiner(2000),Maksym(2010),Maksym(2014),Cenko(2012),Donato(2014),KhabibullinSazonov(2014),Lin(2015)}, known as X-ray bright TDEs \citep{Auchettl(2017)}.  More sophisticated models of compact accretion disks which form soon after the TDE have been able to successfully reproduce many features of X-ray bright TDEs \citep[e.g.][]{StrubbeQuataert(2009),LodatoRossi(2011),MummeryBalbus(2020),Jonker(2020)}.  However, these compact TDE disk models fail to explain the properties of optically or UltraViolet (UV) bright TDEs \citep[e.g.][]{Gezari(2006),Gezari(2008),Gezari(2009),Komossa(2008),vanVelzen(2011),Wang(2011),Wang(2012),Arcavi(2014)}, which also have Eddington luminosities, but have much larger photosphere radii ($\sim 10^{14}-10^{15} \, {\rm cm}$) with lower effective temperatures ($\sim 2-3 \times {10^4} \, {\rm K}$).

One explanation for the low effective temperatures is that the TDE thermal emission does not originate from the accretion disk, but from an outflow supported by radiation pressure from the disk's super-Eddington Accretion rate \citep[e.g.][]{LoebUlmer(1997),StrubbeQuataert(2009),LodatoRossi(2011),MetzgerStone(2016),Roth(2016),CurdNarayan(2019)}.  The TDE's high photosphere radius and low temperature can then be explained by the increased emitting area from the optically thick, expanding outflow, launched from the accretion disk or SMBH.  These outflows can explain the nearly constant temperatures inferred from the spectrum of optically bright TDEs \citep[][]{StrubbeQuataert(2009),Miller(2015)}, and may lead to observable emission or absorption line features in the TDE's spectrum \citep{StrubbeQuataert(2011),Roth(2016),RothKasen(2018)}.  Hydrodynamical simulations show that the outflow can be supported not only by radiation pressure from the compact disk \citep{Dai(2018),CurdNarayan(2019)}, but also by shocks driven by stream-stream collisions during the circularization of stellar debris \citep{LuBonnerot(2020),Liptai(2019)}.  However, to power the outflow, a significant fraction of the tidally disrupted star's rest-mass energy must be liberated ($0.05 \, \Msun \, c^2 \sim 10^{53} \, {\rm erg}$), much larger than the typical energy liberated by an optically bright TDE's early emission ($\sim 10^{49}-10^{51} \, {\rm erg}$, e.g. \citealt{Komossa(2015),vanVelzen(2020)}).  This  so-called missing energy problem has a number of proposed solutions.  {For instance, some argue} most of the rest-mass energy is radiated in the unobservable far UV wavelength bands \citep[e.g.][]{LuKumar(2018),Jonker(2020)}, {while others propose this energy is} carried away by a jet whose emission is unobservable for most TDE viewing angles \citep{Dai(2018)}. {Some authors suggest this energy is never emitted in the first place, but rather becomes trapped} due to the {TDE} disk and outflow's high optical depth (photon trapping; e.g. \citealt{CurdNarayan(2019)}).  The wind model has yet to conclusively address the missing energy problem.

Contending the outflow model for optical TDE emission is the stream-stream collisions model of \cite{Shiokawa(2015),Piran(2015),Krolik(2016)}.  Instead of optical emission originating from an optically thick outflow, it instead comes from the accretion disk, whose annular extent remains extended ($\sim 10^{14} - 10^{15} \, {\rm cm}$) due to inefficient circularization of the debris stream.  The emission itself is powered by shock-heating from stream-stream collisions of disk gas during formation.  A prediction from this model is the broadening of H$\alpha$ and optical emission lines from non-Keplerian motion within the eccentric disk \citep{Piran(2015),Liu(2017),Cao(2018)}.  This model also avoids the problem of how the disk efficiently circularizes, since hydrodynamical simulations show the TDE disk remains extended and eccentric long after formation, because not enough orbital energy is dissipated to drive the disk's eccentricity to zero \citep{Guillochon(2014),Shiokawa(2015),Sadowski(2016),BonnerotLu(2019)}.  After the optical emission is driven by shocks from the circularization process, \cite{Svirski(2017)} argued that X-ray thermal emission from viscous heating of the disk's inner edge may be avoided if the disk remains highly eccentric.  If an elliptical disk annulus loses sufficient angular momentum, its eccentricity may increase, allowing the disk material to cross the SMBH's effective potential barrier and plunge directly into the SMBH without losing orbital energy via viscous dissipation.  But although stream-stream collisions are an intriguing alternative to the wind model to generate TDE optical emission, no detailed model has been put forth to calculate in detail the thermal emission from a poorly circularized TDE disk.

The goal of this work is to construct a highly idealized model of a poorly circularized TDE disk, to gain insight on the disk's structure and dynamical evolution, and to self-consistently calculate the thermal emission from an eccentric TDE disk. {We assume the disk is {inviscid and adiabatic,} and neglect angular momentum transport within the disk through various processes.} Section~\ref{sec:TDEDiskStructureDynamics} presents our model for the poorly circularized TDE disk, and calculates the simplest solutions for a highly eccentric, apsidally aligned, and uniformly precessing hydrodynamical disk.  Section~\ref{sec:TDE_Emit} introduces our model for the TDE thermal emission, and calculates the TDE disk's spectrum.  Section~\ref{sec:Discuss} discusses the theoretical uncertainties and observational implications of our work.  Section~\ref{sec:Conclude} draws our main conclusions.

\section{Eccentric TDE Disk Structure and Dynamics}
\label{sec:TDEDiskStructureDynamics}

When a star of mass $\Ms$ and radius $\Rs$ on a nearly parabolic orbit with periapsis distance $\rp$ from the SMBH of mass $\Mh$ satisfies
\be
\rp \lesssim \Rt = \Rs (\Mh/\Ms)^{1/3},
\ee
the star tidally disrupts.  The stellar debris after disruption bound to the SMBH has a spread in {specific} orbital energy (assuming $\Rs \ll \Rt$)
\be
\Dg E \simeq \frac{G \Mh \Rs}{\Rt^2}.
\ee
Throughout the rest of this work, we define the dimensionless stellar mass $\bMs$, radius $\bRs$, and SMBH mass $\bMh$ via
\be
\bMs = \frac{\Ms}{1 \, \Msun},
\hspace{3mm}
\bRs = \frac{\Rs}{1 \, \Rsun},
\hspace{3mm}
\bMh = \frac{\Mh}{10^6 \, \Msun},
\label{eq:norm}
\ee
normalized to typical TDE values \citep[e.g.][]{vanVelzen(2020)}.

After disruption, the debris is expected to re-accrete, and form a disk orbiting the SMBH.  The shortest orbital period within the debris stream is
\begin{align}
&\tfo = \frac{2 \pi G \Mh}{(2 \Dg E)^{3/2}} = \frac{\pi \Rt^3}{\sqrt{2 G \Mh \Rs^3}}
= 41 \frac{\bMh^{1/2}\bRs^{3/2}}{\bMs} \ {\rm days},
\label{eq:tfo}
\end{align}
The specific orbital energy of debris which re-accretes onto the TDE disk is
\be
\Ef(t) = -\frac{(2\pi G \Mh)^{2/3}}{2 t^{2/3}} = -\Dg E \left( \frac{\tfo}{t} \right)^{2/3},
\ee
where we will assume from now on that debris with orbital period $\tf$ accretes onto the disk at time $t = \tf$.  The accretion rate of fallback material onto the TDE disk is then \citep{Rees(1988),EvansKochanek(1989)}
\begin{align}
\frac{\der M_{\rm f}}{\der t} &= \frac{\der M_{\rm f}}{\der E_{\rm f}} \frac{\der E_{\rm f}}{\der t} = \frac{\Ms}{3 \tfo} \left( \frac{\tfo}{t} \right)^{5/3}
\nonumber \\
&= 134 \, \dot M_{\rm Edd} \left( \frac{\eta}{0.1} \right) \frac{\bMs}{\bMh^{3/2} \bRs^{3/2}}  \left( \frac{\tfo}{t} \right)^{5/3}
\label{eq:dMfdt}
\end{align}
where $\dot M_{\rm Edd} = 4\pi G \Mh/(\eta c \kg)$ is the Eddington accretion rate and $\eta$ is the accretion efficiency factor.  Here, we have assumed the opacity $\kg$ is dominated by electron-scattering ($\kg = 0.34 \ {\rm cm}^2/{\rm g}$), and distribution of binding energy in the star was constant ($\der M_{\rm f}/\der \Ef \approx {\rm constant}$), an assumption which may break down at early times ($t \lesssim {\rm few} \ \tfo$) due to the realistic structure of a star \citep[e.g.][]{Lodato(2009),GuillochonRamirez-Ruiz(2013),Law-Smith(2019),Golightly(2019),Ryu(2020a),Ryu(2020b)}, and at late times ($t \gtrsim {\rm few} \ \tfo$) due to debris self-gravity \citep{CoughlinNixon(2019)} or non-parabolic stellar orbits \citep{Hayasaki(2018),ParkHayasaki(2020)}.  The semi-major axis of stellar debris which accretes onto the TDE disk is
\begin{align}
&\af(t) = -\frac{G \Mh}{2 \Ef} = \frac{\Rt^2}{2 \Rs} \left( \frac{t}{\tfo} \right)^{2/3}
\nonumber \\
&= 2356 \frac{\bRs}{\bMh^{1/3} \bMs^{2/3}}  \left( \frac{t}{\tfo} \right)^{2/3} \Rg,
\label{eq:af}
\end{align}
where $\Rg = G \Mh/c^2$ is the gravitational radius.  Assuming that all stellar debris has the same periapsis distance to the SMHB (since $\Rs \ll \Rt$), the eccentricity of the accreted fallback debris with time is
\begin{align}
&\ef(t) = 1 - \frac{\rp}{\af} = 1 - \frac{2 \rp \Rs}{\Rt^2} \left( \frac{\tfo}{t} \right)^{2/3}
\nonumber \\
&= 1 - \frac{0.02}{\bg} \frac{\bMs^{1/3}}{\bMh^{1/3}} \left( \frac{\tfo}{t} \right)^{2/3},
\label{eq:ef}
\end{align}
{where $\bg = \Rt/\rp$ is the dimensionless impact parameter.  {We keep the parameter $\beta$ in all analytical expressions for generality, but set $\beta=1$ for all numerical calculations.}  This assumption only affects the eccentricity of the fall-back debris, and has no {effect} on the surface density or internal energy of our TDE disk model.}  

In the subsequent subsections, we will build our model of a poorly circularized disk (Sec.~\ref{sec:TDEDiskStructure}), introduce the secular formalism which we will use to calculate the structure and dynamical evolution of highly eccentric TDE disks (Sec.~\ref{sec:EccDiskForm}), and present our results for eccentric TDE disk solutions (Sec.~\ref{sec:EccTDEDiskSols}).

\subsection{Poorly Circularized TDE Disk Model}
\label{sec:TDEDiskStructure} 

{
This section develops our idealized background model for the eccentric TDE disk.  Our model assumes after one orbital period, some fraction of the debris orbital energy is converted into internal energy, heating the disk via the circularization of debris.  The orbital energy lost gives the semi-major axis of the newly-formed disk annulus in relation to the semi-major axis of the original debris stream.  The disk surface density is then computed by integrating the fall-back rate over the disk's annular extent, to compute the disk mass and {its} gradient with distance from the SMBH.  Within this section, we assume the disk annuli have no eccentricity.
{The ``reference circular disk'' model developed in this section is a standard, circular disk in which the mass and entropy are distributed in semimajor axis the same way as a more general eccentric disk.  We discuss how this model generalizes to eccentric disks in following sections.
}
}

Hydrodynamical simulations show stream-stream collisions during the circularization of debris can heat the TDE disk \citep[e.g.][]{Guillochon(2014),Shiokawa(2015)}, but recent simulations give conflicting results on the efficiency of this process \citep[e.g.][]{BonnerotLu(2019),Liptai(2019)}.  We parameterize the uncertainty of TDE circularization by assuming after one orbital period of the stellar debris $\tf$, a fraction $\cV$ of the debris {orbital binding} energy is dissipated (due to self-intersection of elliptical trajectories from apsidal precession, nozzle shocks, Magneto-Rotational Instability, etc., see discussions in e.g. \citealt{Guillochon(2014), Bonnerot(2017)}), which is converted into internal energy within the disk.  Quantitatively, assuming the disk material orbits the SMBH on Keplerian orbits with a range of semi-major axis $a$, the specific orbital energy of disk material is
\be
\Eorb = - \frac{G \Mh}{2 a} = (1+\cV) \Ef = -(1+\cV) \frac{G \Mh}{2 \af}. 
\label{eq:Eorb}
\ee
The inner semi-major axis of the disk is thus
\begin{align}
&\ain = \frac{2356}{1+\cV} \frac{\bRs}{\bMh^{1/3} \bMs^{2/3}} \Rg
\nonumber \\
&= \frac{3.48 \times 10^{14}}{1+\cV} \frac{\bRs \bMh^{2/3}}{\bMs^{2/3}} \, {\rm cm},
\label{eq:ain}
\end{align}
while the outer semi-major axis of the disk is
\be
\aout(t) = \frac{\af(t)}{1+\cV} = \ain \left( \frac{t}{\tfo} \right)^{2/3}.
\label{eq:aout}
\ee
{Although we give explicit expressions for how $\aout$ evolves with time, because hydrodynamical simulations find TDE disks form with $\aout \gg \ain$ very early on ($t \lesssim {\rm few} \, \tfo$, e.g. \citealt{Shiokawa(2015),BonnerotLu(2019),Liptai(2019)}), we leave $\aout/\ain$ as a free parameter.}
The dissipated orbital energy is converted into internal energy within the disk:
\be
\beps^\circ = - \frac{\cV}{1+\cV} \Eorb,
\label{eq:epsInt}
\ee
where $\beps^\circ$ is the specific internal energy for a circular disk.  Notice that $\Eorb + \beps^\circ = \Ef$, as required by energy conservation.  Throughout this work, we denote by $X^\circ$ disk properties $X$ for a reference circular and axisymmetric disk.

Using equation~\eqref{eq:dMfdt}, we can calculate the total disk mass with time:
\be
\Md(t) = \int_{\tfo}^t \frac{\der \Mf}{\der t'} \der t' = \frac{\Ms}{2} \left[ 1 - \left( \frac{\tfo}{t} \right)^{2/3} \right].
\label{eq:Md}
\ee
Since fallback material with orbital period $\tf$ is accreted at time $t = \tf$, the total disk mass contained within $a$ is
\be
M(a) = \frac{\Ms}{2} \left( 1 - \frac{\ain}{a} \right),
\label{eq:M}
\ee
hence the disk mass radial gradient is
\be
M_a \equiv \frac{\der M}{\der a} = \frac{\Ms \ain}{2 a^2}.
\label{eq:Ma}
\ee
{Equation~\eqref{eq:Ma} is a direct consequence of assuming the mass distribution of stellar debris with binding energy is constant ($\der M_{\rm f}/\der \Ef = {\rm constant}$), and deviations of $\dotMfb$ from $\dotMfb \propto t^{-5/3}$ will affect the mass distribution $M_a$.}

With $\beps^\circ$ and $M_a$ specified, we may calculate other properties of the disk.  The surface density of a circular disk $\Sg^\circ$ is related to $M_a$ via
\be
\Sg^\circ = \frac{M_a}{2\pi a} = \frac{\Ms}{4\pi \ain^2 \ta^3},
\label{eq:Sg}
\ee
where $\ta = a/\ain$ is the re-scaled semi-major axis.  The vertically integrated pressure for a circular disk $P^\circ$ with an adiabatic equation of state is related to $\beps^\circ$ and $\Sg^\circ$ by
\be
P^\circ = (\cg-1)\Sg^\circ \beps^\circ = \frac{(\cg-1)\cV}{8\pi (1+\cV)} \frac{G \Mh \Ms}{\ain^3 \ta^4}.
\ee
The circular disk scale-height $H^\circ$ is then
\be
\frac{H^\circ}{a} = \sqrt{ \frac{P^\circ}{\Sg^\circ n^2 a^2} } = \sqrt{ \frac{(\cg-1)\cV}{1 + \cV} },
\label{eq:H}
\ee
where $n = \sqrt{G \Mh/a^3}$ is the mean-motion.  Notice $H^\circ/a \sim 1$ when $\cV\gtrsim 1$, but $H^\circ/a \ll 1$ when $\cV \ll 1$.

Assuming the disk is dominated by radiation pressure ($\cg = 4/3$), the disk's circular mid-plane temperature is given by
\begin{align}
    &T_0^\circ = \left( \frac{3 P^\circ}{\ard H^\circ} \right)^{1/4}
    \nonumber \\
    &= 1.13 \times {10^5} \, \cV^{1/8} (1+\cV)^{7/8}
    \frac{\bMs^{11/12}}{ \bRs \bMh^{5/12}} \frac{1}{\ta^{5/4}}  \ {\rm K},
    \label{eq:T0}
\end{align}
where $\ard$ is the radiation constant.  Since the ratio of gas to radiation pressure in the disk's midplane is (for a circular disk)
\begin{align}
    &\frac{p_{\rm g}}{p_{\rm rd}} = \frac{k_{\rm B} T_0^\circ \Sg^\circ/(\mu_{\rm e} m_{\rm p} H^\circ)}{\ard {T_0^\circ}^4/3}
    \nonumber \\
    &\simeq 2.38 \times 10^{-4} \left( \frac{1+\cV}{\cV} \right)^{7/8} \frac{\bMs^{1/4}}{\bMh^{3/4}} \frac{1}{\ta^{1/4}},
    \label{eq:gas_rad_press}
\end{align}
where $k_{\rm B}$ is the Boltzmann constant, $\mu_{\rm e} = 0.62$ is the mean molecular weight for an ionized gas of solar composition, and $m_{\rm p}$ is the proton mass, we see the assumption of a radiation-pressure dominated disk is justified.  The disk's optical depth is (for a circular disk)
\be
\tau^\circ = \kg \Sg^\circ = 445 \, (1+\cV)^2 \frac{\bMs^{7/3}}{\bRs^2 \bMh^{4/3}} \frac{1}{\ta^3}.
\label{eq:tau}
\ee
Only when the disk's semi-major axis satisfies $a \gtrsim 8 \, \ain$ for standard TDE disk parameters does the disk become optically thin ($\tau < 1$).  Because the TDE disk models considered in this work are confined to a relatively narrow range in semi-major axis ($\aout \lesssim {\rm few} \, \ain$), the disk is optically thick ($\tau \ge 1$).

We emphasize that our background disk model idealizes the highly complex circularization process the stellar debris is expected to undergo.  Simulations find that the disk which forms soon after the TDE is not coherent, but a mess of colliding streamlines on elliptical trajectories \citep[e.g.][]{Shiokawa(2015),Sadowski(2016),BonnerotLu(2019),LuBonnerot(2020)}.  Therefore, the resulting eccentricity profiles calculated in this work should be taken with a grain of salt.

{

\subsection{Sources of Energy Dissipation}
\label{sec:EnergyDiss}

\begin{figure}
	\includegraphics[width=\columnwidth]{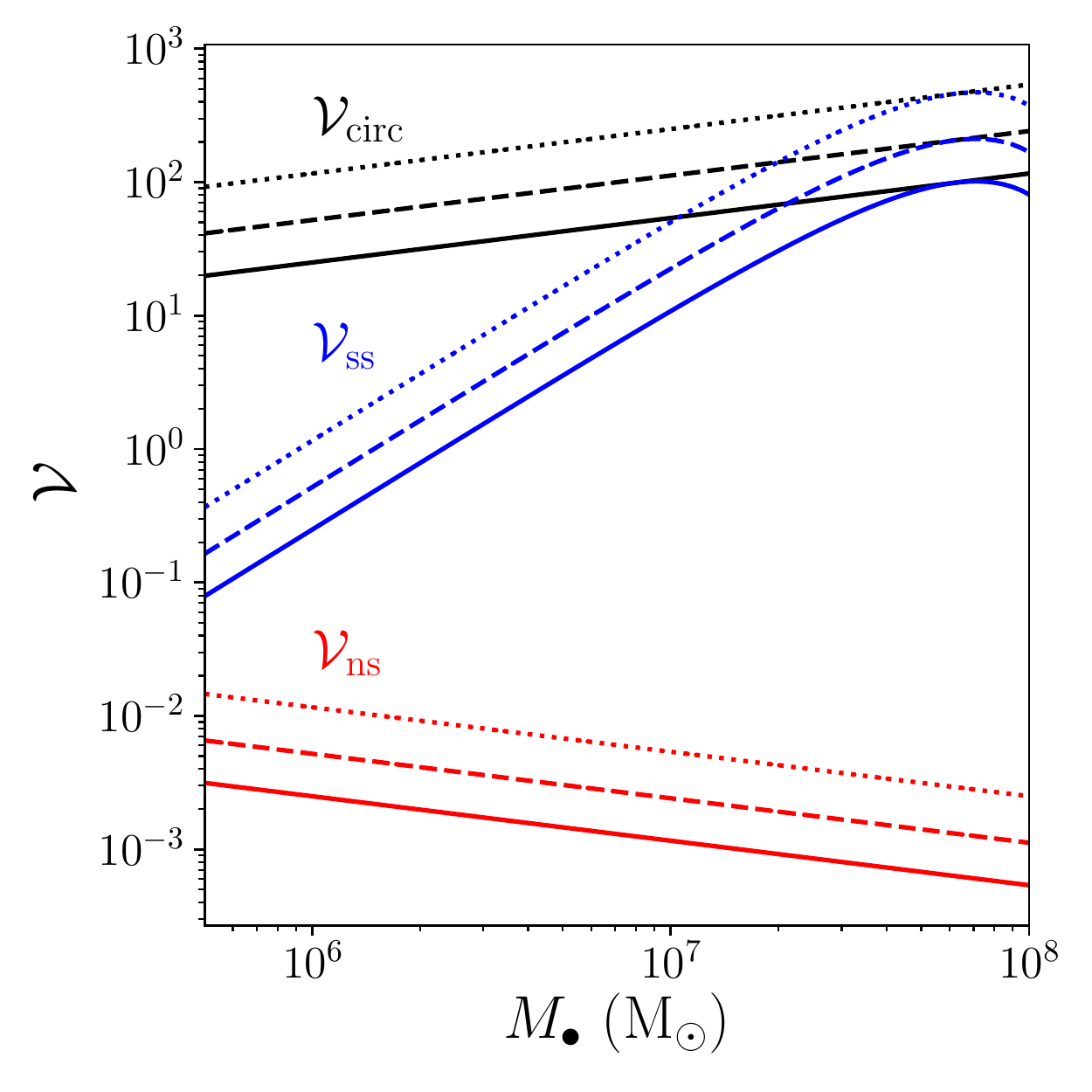}
    \caption{
    {
    Fraction of orbital energy dissipated after one orbital period $\cV$ due to the nozzle-shock near pericenter $\cV_{\rm ns}$ (red lines, eq.~\ref{eq:cVns}) and stream-stream collisions near apocenter $\cV_{\rm ss}$ (blue lines, eq.~\ref{eq:cVss}), compared with the fraction of orbital energy which needs to be dissipated to circularize the orbit $\cV_{\rm circ}$ (black lines, eq.~\ref{eq:cVcirc}), as a function of the SMBH mass $\Mh$.  Different line styles denote different times, with $t = 1 \, \tfo$ (solid lines), $t = 3 \, \tfo$ (dashed lines), and $t = 10 \, \tfo$ (dotted lines).
    }
    }
    \label{fig:cV_plot}
\end{figure}

 Our treatment of the circularization process parameterizes the fraction of debris orbital energy lost after an orbital period via $\cV$ (see eq.~\ref{eq:Eorb}).  This section reviews estimates for the two primary sources which have been argued to be the most efficient ways orbital energy is dissipated during the circularization of the debris: the nozzle shock near pericenter from adiabatic compression, and stream-stream collisions near apocenter from apsidal precession (see also e.g. \citealt{Krolik(2020)}).  To { estimate the upper bound on the energy dissipated} through these shocks, we make the assumption the collision between fluid streams is inelastic.  Specifically, we assume the ingoing and outgoing streams have a similar density, so that momentum conservation requires the {normal} components of the ingoing stream velocity (relative to the outgoing stream velocity) be dissipated during the shock.
 
 We first review nozzle-shock dissipation.  Assuming the stellar debris {lies} on ballistic orbits after the star is disrupted at the tidal radius $\Rt$, angular momentum conservation dictates the orbital planes of the debris will be confined within the angle $\theta \simeq \Rs/\sqrt{\rp \Rt}$ \citep{CarterLuminet(1982)}.  Hence the {normal} velocity component at pericenter of the debris orbit is $v_z \sim \theta v_{\rm p}$, where $v_{\rm p} = \sqrt{G \Mh/\rp}$ is the orbital velocity at pericenter.  One can show after an orbital period of the debris, adiabatic compression with a $\cg = 5/3$ equation of state (a reasonable assumpiton, since the cold debris {is} dominated by gas pressure) implies the maximum {normal} velocity at pericenter $v_z$ remains unchanged \citep{Guillochon(2014)}.  Hence the energy dissipated via the nozzle shock is of order
 \be
 \dg E_{\rm ns} \sim v_z^2 = \bg^2 \left( \frac{G \Ms}{\Rs} \right).
 \label{eq:dEns}
 \ee
 Relating $\dg E_{\rm ns}$ to the fractional energy lost after an orbital period $\cV_{\rm ns} = \dg E_{\rm ns}/\Ef$, we have
 \be
 \cV_{\rm ns} \sim \frac{\bg^2}{2} \left( \frac{\Ms}{\Mh} \right) \left( \frac{\af}{\Rs} \right) = 2.5 \times 10^{-3} \, \bg^2 \frac{\bMh^{1/3}}{\bMh^{1/3}} \left( \frac{t}{\tfo} \right)^{2/3}.
 \label{eq:cVns}
 \ee
 However, the actual amount of dissipation experienced by the stellar debris may depend sensitively on the fluid's viscosity, since an adiabatic gas can rebound after pericenter passage with little dissipation \citep{LynchOgilvie(2020)}
 
 The second significant source of energy dissipation is stream-stream collisions near apocenter from GR apsidal precssion.  As in \cite{Dai(2015),Bonnerot(2017)}, we calculate the ingoing collision stream velocities by finding the intersection point of two elliptical orbits, shifted by an instantaneous pericenter shift $\dg \pom$ from GR apsidal precession.  Since an eccentric orbit undergoes apsidal precession at a rate (assuming $\af \gg \Rg$)
 \be
\left. \frac{\pd \pom}{\pd t} \right|_{\rm GR} = \frac{3 \Rg n}{a (1-e^2)},
\label{eq:dpomdt_GR}
\ee
we have after one orbital period
\be
\dg \pom \simeq \frac{3 \pi \Rg}{\rp} = 11.5^\circ \, \bg \frac{\bMh^{2/3} \bMs^{1/3}}{\bRs}.
 \label{eq:dpom}
\ee
With $\dg \pom$, one can calculate the relative stream velocities during the collision, leading to a loss of energy of \citep{Dai(2015),Bonnerot(2017)}
\begin{align}
\dg E_{\rm ss} &= \frac{\ef^2 G \Mh}{2\af (1-\ef^2)} \sin^2 \left( \frac{\dg \pom}{2} \right)
\simeq \frac{G \Mh}{4 \rp} \sin^2 \left( \frac{\dg \pom}{2} \right),
\label{eq:dEss}
\end{align}
hence
\be
\cV_{\rm ss} = \frac{\dg E_{\rm ss}}{\Ef} = \frac{\af}{2 \rp} \sin^2 \left( \frac{\dg \pom}{2} \right).
\label{eq:cVss}
\ee
When $\dg \pom \ll 1$, equation~\eqref{eq:cVss} reduces to
\be
\cV_{\rm ss} \simeq \frac{9 \pi^2}{16} \frac{\Rg^2 \af}{\rp^3} = 0.125 \, \bg^3 \frac{\bMh^{5/3} \bMs^{1/3}}{\bRs^2} \left( \frac{t}{\tfo} \right)^{2/3}.
\label{eq:cVss_approx}
\ee

Different sources of energy dissipation need to be compared with the total loss of energy required to completely circularize the disk.  If the debris circularizes with constant orbital angular momentum, the final semi-major axis is given by $a_{\rm circ} = (1 - \ef^2)\af \simeq 2 \rp$.  Hence the orbital energy of a completely circularized debris stream is given by
\be
E_{\rm circ} = - \frac{G \Mh}{4 \rp},
\label{eq:Ecirc}
\ee
so the fractional binding energy which must be lost to circularize the stream is
\be
\cV_{\rm circ} = \frac{E_{\rm circ}}{\Ef} = \frac{\Rt^2}{4 \rp \Rs} \left( \frac{t}{\tfo} \right)^{2/3} = \frac{25}{\bg} \frac{\bMh^{1/3}}{\bMs^{1/3}} \left( \frac{t}{\tfo} \right)^{2/3}.
\label{eq:cVcirc}
\ee
We see that $\cV_{\rm circ} \gg 1$ for typical parameters.

Figure~\ref{fig:cV_plot} plots $\cV_{\rm ns}$ and $\cV_{\rm ss}$, compared with the orbital energy which must be dissipated to circularize the debris orbit $\cV_{\rm circ}$ (eq.~\ref{eq:cVcirc}).  We see energy dissipation from stream-stream collisions dominates over that from the nozzle-shock ($\cV_{\rm ss} \gg \cV_{\rm ns}$) over the SMBH range relevant for TDEs ($10^6 \ \Msun \lesssim \Mh \lesssim 10^8 \ \Msun$).  As the TDE evolves in time, the energy dissipated via shocks is expected to increase, but only by factors of order unity over much of the TDE disk's lifetime ($t \lesssim 10 \, \tfo$).  Only when $\Mh \gtrsim 1-3 \times 10^7 \ \Msun$ can enough orbital energy be dissipated to completely circularize the orbit ($\cV_{\rm ss} \gtrsim \cV_{\rm circ}$).  We also see why other simulations find $\cV_{\rm ns} \gg \cV_{\rm ss}$ at early times, since they used SMBH mass values much lower than those typical of TDEs ($\Mh \ll 10^6 \, \Msun$) due to computational constraints \citep[e.g.][]{Shiokawa(2015)}, and $\cV_{\rm ns} \propto \Mh^{-1/3}$ while $\cV_{\rm ss} \propto \Mh^{5/3}$ when $\Mh \lesssim 10^7 \, \Msun$.

From this, we see for the SMBH mass range investigated in this work ($10^6 \ \Msun \lesssim \Mh \lesssim 10^7 \ \Msun$), stream-stream collisions dominate the energy dissipation during the circularization process, causing $\cV$ to lie within the range $0.1 \lesssim \cV \lesssim 10$.  Although the energy dissipated via shocks is likely highly {non-axisymmetric} with $\cV$ evolving in time, because our disk model is highly idealized, we assume $\cV$ to be {axisymmetric} and constant in time for simplicity.  Deviations from this assumption will affect the thermal structure of the TDE disk.  Also, we assume a constant fraction $\cV$ of the debris {orbital binding energy $|\Ef|$}, rather than the debris kinetic energy $\frac{1}{2} v^2$, is converted into internal energy $\beps$.  This is for simplicity: a more detailed treatment would have to make specific assumptions about the process which dissipates the {debris} orbital energy, and would affect the orientation of the newly-formed eccentric disk.

Within this work, we assume the TDE disk forms when $t \ge \tfo$, but in reality this formation time depends on where most of the debris orbital energy is dissipated.  If most dissipation occurs near pericenter (via $\cV_{\rm ns}$), the disk will form when $t \gtrsim 1.5 \, \tfo$, while dissipation near apocenter (via $\cV_{\rm ss}$) will have disk formation occur when $t \gtrsim 2 \, \tfo$.

}

\subsection{Eccentric Disk Formalism}
\label{sec:EccDiskForm}

\begin{figure}
	\includegraphics[width=\columnwidth]{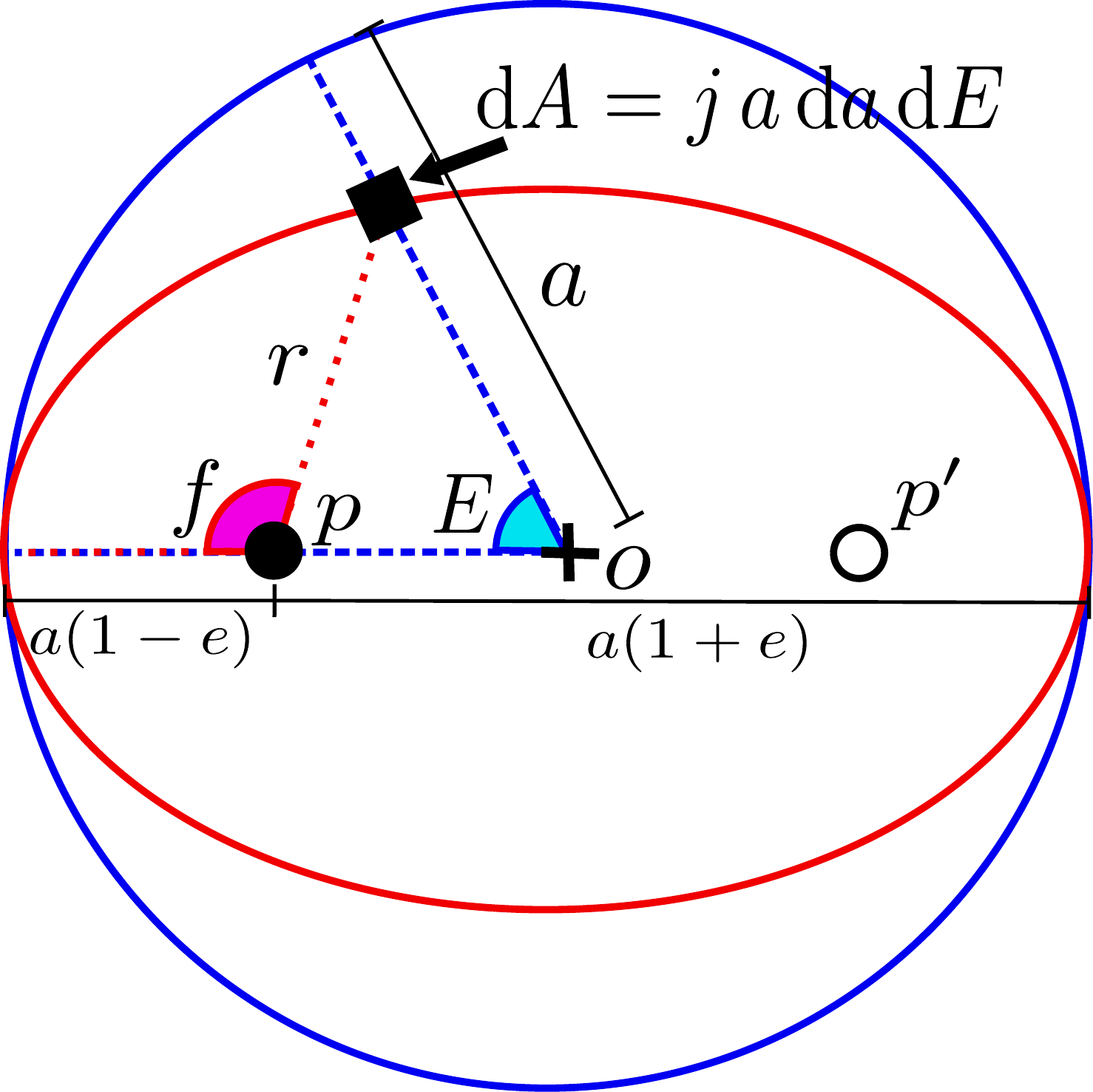}
	\includegraphics[width=\columnwidth]{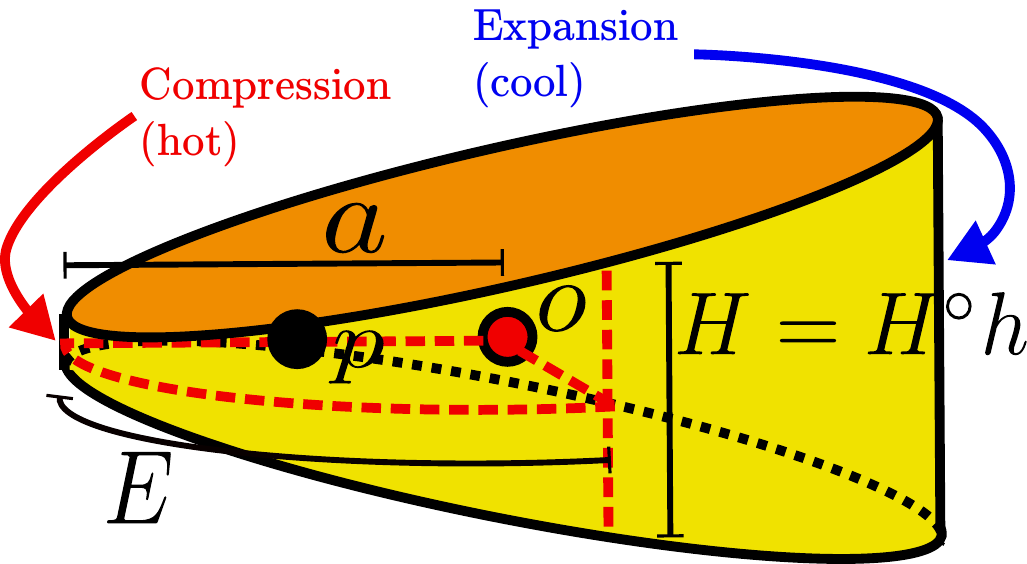}
    \caption{
    \textit{Top diagram:} Coordinate system for a massless fluid element (black square) on an orbit with eccentricity $e$ (red ellipse) surrounded by a circumscribed circle ({reference circular disk,} blue), where $o$ is the origin, $p$ is one focus point (location of central mass), while $p'$ is another focus point.  Other quantities are distance of fluid element from central mass $r$, true anomaly $f$, semi-major axis $a$, and eccentric anomaly $E$.  The infinitesimal area of a fluid element at $(a,E)$ is $\der A = j a \, \der a \, \der E$, where $j$ is the (dimensionless) Jacobian (eq.~\ref{eq:j}).  \textit{Bottom Diagram:} Vertical scale-height $H(a,E) = H^\circ(a) h(E)$ for an eccentric annulus orbiting a central mass at $p$.  Because $h$ is small (large) near periapsis (apoapsis), the fluid's temperature is high (low) when $E\approx 0$ ($E \approx \pi$).
    }
    \label{fig:Coos}
\end{figure}

\begin{figure}
	\includegraphics[width=\columnwidth]{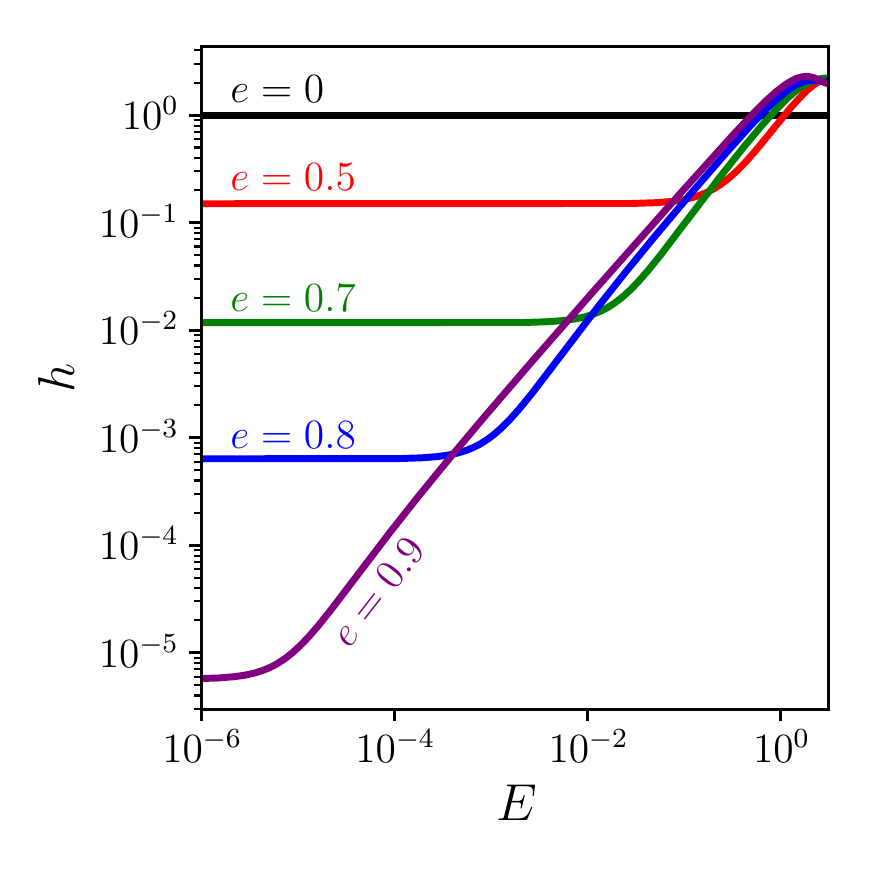}
    \caption{
    Dimensionless scale-height $h$ as a function of eccentric anomaly $E$ (see Fig.~\ref{fig:Coos}), for the eccentricity values $e$ indicated.  Here, $a e_a = 0$ and $a \pom_a = 0$.
    }
    \label{fig:h(E)}
\end{figure}

\begin{figure}
	\includegraphics[width=\columnwidth]{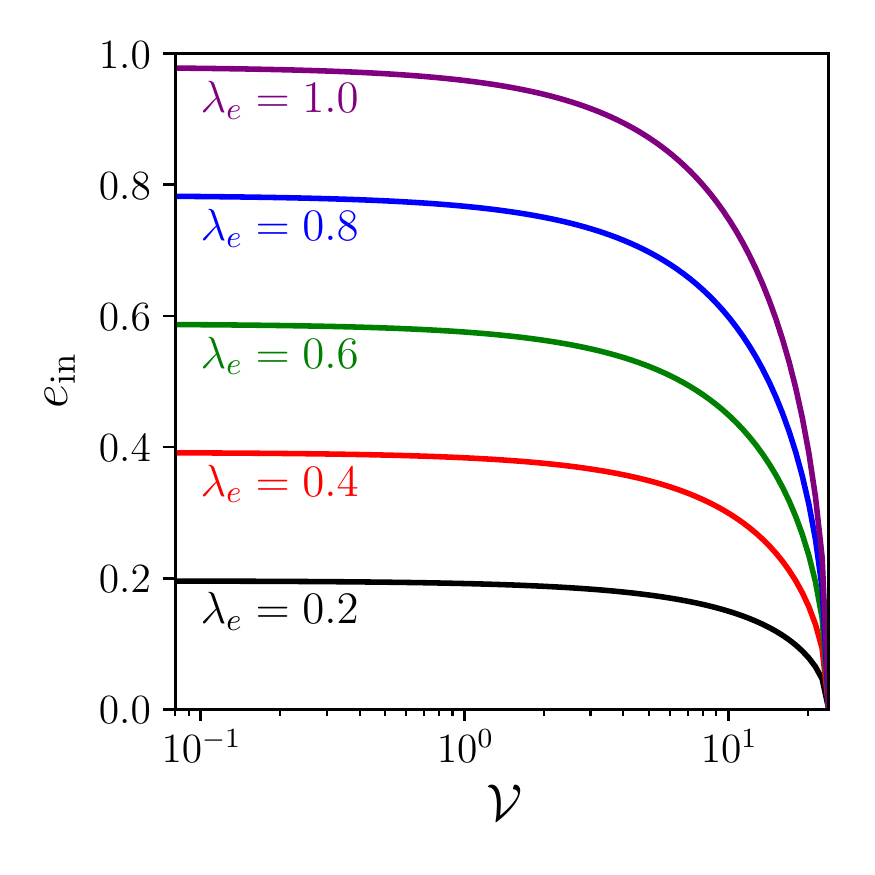}
    \caption{
    Inner disk's eccentricity $e_{\rm in} = e(\ain)$ (eq.~\ref{eq:ein}) as a function of the fallback debris orbital energy dissipation parameter $\cV$ (see eqs.~\ref{eq:Eorb} \&~\ref{eq:epsInt}), for the $\lam_e$ values indicated.  Here, $\Mh = 10^6 \, \Msun$.
    }
    \label{fig:ein}
\end{figure}

To calculate the eccentricity profile and dynamical evolution of the TDE disk, we use the secular, adiabatic formalism of \cite{OgilvieLynch(2019)}, taking into account the vertical structure of the disk gas as it varies over an elliptical orbit (see also \citealt{Ogilvie(2001),OgilvieBarker(2014)}).  The top diagram of Figure~\ref{fig:Coos} displays the coordinate system of \cite{OgilvieLynch(2019)}, describing a fluid element along an eccentric orbit with eccentricity $e$ and longitude of periapsis $\pom$, whose location can be completely specified by $a$ and the eccentric anomaly $E$ (see also \citealt{MurrayDermott(1999)}).  Notice the infinitesimal area $\der A$ of a fluid element at $(a,E)$ is given by $\der A = j a \, \der a \, \der E$, where
\be
j = \frac{1 - e(e + a e_a) - a e_a \cos E}{\sqrt{1-e^2}} - a e \pom_a \sin E
\label{eq:j}
\ee
is the dimensionless Jacobian of the coordinate system, and we will define $X_a \equiv \pd X/\pd a$ for all functions $X$.  Mass continuity then gives how the surface density varies along an elliptical orbit, $\Sg = \Sg^\circ/j$, so $\tau = \tau^\circ/j$.  We note that both $\Sg$ and $\tau$ typically vary by factors of order unity when compared with their reference circular $\Sg^\circ$ and $\tau^\circ$ values: even for the highly eccentric case of $e = -a e_a = 0.9$ (with $\pom_a = 0$), $j$ varies from a minimum value of 0.23 to a maximum value of 4.35.  Hence even highly eccentric disks are optically thick ($\tau \ge 1$).

\cite{OgilvieLynch(2019)} model an eccentric disk as a continuous nested sequence of elliptical orbits, which communicate with one another through pressure.  The secular (orbit averaged) evolutionary equations for the disk eccentricity magnitude $e = e(a,t)$ and twist $\pom = \pom(a,t)$ are given by
\begin{align}
M_a \frac{\pd e}{\pd t} &= \frac{\sqrt{1-e^2}}{n a^2 e} \left[ \frac{\pd \cH_a}{\pd \pom} - \frac{\pd}{\pd a} \left( \frac{\pd \cH_a}{\pd \pom_a} \right) \right] + M_a \left. \frac{\pd e}{\pd t} \right|_{\rm ext},
\label{eq:dedt} \\
M_a \frac{\pd \pom}{\pd t} &= - \frac{\sqrt{1-e^2}}{n a^2 e} \left[ \frac{\pd \cH_a}{\pd e} - \frac{\pd}{\pd a} \left( \frac{\pd \cH_a}{\pd e_a} \right) \right] + M_a \left. \frac{\pd \pom}{\pd t} \right|_{\rm ext},
\label{eq:dpomdt}
\end{align}
where $\cH_a = \cH_a^\circ F$ is the Hamiltonian density, $\cH_a^\circ = 2\pi a P^\circ = (\cg-1) M_a \beps^\circ$ is the Hamiltonian density for a 2D circular disk, and
\be
F = \frac{\cg+1}{4\pi (\cg-1)} \int_0^{2\pi} \frac{1 - e \cos E}{(jh)^{\cg-1}} \der E
\label{eq:F}
\ee
is the dimensionless geometric part of the Hamiltonian for a three-dimensional disk with adiabatic index $\cg$ (we take $\cg = 4/3$), and $h$ is the dimensionless scaleheight defined below.
{Physically, $\cH_a$ is the total energy {(excluding orbital energy)} per unit semi-major axis $a$ of an eccentric fluid annulus, averaged over an orbital period.  The function $F$ is the geometric part of the orbit-averaged energy density, which only depends on $e$ and $a e_a$.}
Notice when $e = a e_a = 0$, $F = (\cg+1)/[2(\cg-1)]$.

The disk scale-height can be written as $H = h H^\circ$, where the dimensionless function $h = h(E)$ specifies how disk thickness varies over an eccentric orbit, whose azimuthal dependence is determined by the balance of gravity with pressure \citep{Ogilvie(2001),OgilvieBarker(2014),OgilvieLynch(2019)}:
\be
(1-e \cos E) \frac{\der^2 h}{\der E^2} - e \sin E \frac{\der h}{\der E} + h = \frac{(1-e \cos E)^3}{j^{\cg-1} h^\cg}.
\label{eq:h}
\ee
Since $h$ is an even function periodic over the interval $E \in [-\pi,\pi]$, $h(E)$ is calculated numerically using the boundary conditions $h'(0) = h'(\pi) = 0$ with the shooting method \citep{Press(2002)}.  The bottom diagram of Figure~\ref{fig:Coos} displays a cartoon of how $H$ varies along an eccentric orbit, and because the fluid is adiabatic, is compressionally heated near periapsis.  Figure~\ref{fig:h(E)} displays solutions for $h(E)$ for different $e$ values.  The function $h$ has a minimum near periapsis ($E = 0$), and typically has a maximum near apoapsis ($E \approx \pi$).  When $e$ is increased, the value of $h$ near periapsis can be reduced below unity by many orders of magnitude.  The equation of state for an adiabatic gas then gives the vertically integrated pressure $P = P^\circ j^{-\cg} h^{-(\cg-1)}$, and midplane temperature $T_0 = T_0^\circ (j h)^{-\cg/4}$.

In this work, we are interested in a particular class of solutions, where $\pom$ takes the simple form
\be
\pom = \om t + \pom_0,
\label{eq:om}
\ee
where $\pom_0 = \pom|_{t=0}$ is the initial condition, and $\om$ is the constant, global precession frequency of the eccentric disk.  Physically, one can have an untwisted eccentric disk ($\pom_a=0$, eq.~\ref{eq:om}) when the inertial force from $\om$ balances the twisting forces from pressure.  Then $\pd e/\pd t = 0$ (assuming $\pd e/\pd t|_{\rm ext} = 0$; eq.~\ref{eq:dedt}), while equation~\eqref{eq:dpomdt} reduces to
\begin{align}
M_a \om = - \frac{\sqrt{1-e^2}}{n a^2 e} \cH_a^\circ &\left[ \frac{\pd F}{\pd e} - a e_a \frac{\pd^2 F}{\pd e \pd f} - (2 a e_a + a^2 e_{aa}) \frac{\pd^2 F}{\pd f^2} \right.
\nonumber \\
& \left. - \frac{\der \ln \cH_a^\circ}{\der \ln a} \frac{\pd F}{\pd f} \right] + M_a \left. \frac{\pd \pom}{\pd t} \right|_{\rm GR},
\label{eq:dpomdt_red}
\end{align}
where $f = e + a e_a$ (not to be confused with the true anomaly $f$, see Fig.~\ref{fig:Coos}){, while $\pd \pom/\pd t|_{\rm GR}$ is given by equation~\eqref{eq:dpomdt_GR}.}  Equation~\eqref{eq:dpomdt_red} may be re-arranged to
\begin{align}
-\tom \ta^{3/2} \frac{e}{\sqrt{1-e^2}} = \ &\frac{\pd F}{\pd e} - a e_a \frac{\pd^2 F}{\pd e \pd f} - (2 a e_a + a^2 e_{aa}) \frac{\pd^2 F}{\pd f^2}
\nonumber \\
&- \frac{\der \ln \cH_a^\circ}{\pd \ln a} \frac{\pd F}{\pd f} - \frac{\dg_{\rm GR}}{\ta} \frac{e}{(1-e^2)^{3/2}},
\label{eq:e(a)}
\end{align}
where
\begin{align}
    \tom &= \om \left. \frac{n a^2}{(\cg-1) \beps^\circ} \right|_{a=\ain},
    \label{eq:tom} \\
    \dg_{\rm GR} &= \left. \frac{n a^2}{(\cg - 1) \beps^\circ} \frac{\pd \pom}{\pd t} \right|_{{\rm GR}, \, e=0, \, a=\ain} = \frac{6 \Rg (1 + \cV)}{(\cg-1) \ain \cV}
    \nonumber \\
    &= 7.64 \times 10^{-3} \,  \frac{ (1+\cV)^2}{\cV} \frac{\bMh^{1/3} \bMs^{2/3}}{\bRs}.
    \label{eq:dg_GR}
\end{align}
Because $\cH_a^\circ \propto M_a \beps^\circ \propto a^{-3}$, we have $\der \ln \cH_a^\circ/\der \ln a = -3$ in our model.
{Physically, $\dg_{\rm GR}$ is the GR apsidal precession rate $\dot \pom_{\rm GR}$ multiplied by the radial communication timescale for an eccentric disturbance $t_{\rm ecc} = n a^2/[(\cg-1)\beps^\circ]$ when $e \ll 1$.}

Because not enough energy has been dissipated to circularize the fallback debris after one orbital period, the newly formed TDE disk will be born with a substantial eccentricity \citep[e.g.][]{Shiokawa(2015),Sadowski(2016),BonnerotLu(2019)}.  Assuming angular momentum conservation, the eccentricity of the most tightly bound debris after one orbital period is (assuming $\rp \ll \af[\tfo]$)
\begin{align}
    &e_{\rm circ} = \sqrt{ 1 - (1+\cV) \left[ 2 - \frac{\rp}{\af(\tfo)} \right] \frac{\rp}{\af(\tfo)} }
    \nonumber \\
    &\simeq \sqrt{ 1 - 4 \bg (1+\cV) \left( \frac{\Ms}{\Mh} \right)^{1/3} }.
    \label{eq:e_circ}
\end{align}
To model the excitation of the disk's eccentricity by the accretion of fallback material, we set the disk eccentricity at $\ain$ to be
\be
\ein = \lam_e e_{\rm circ}.
\label{eq:ein}
\ee
Here $\lam_e \le 1$ is a dimensionless free parameter to model the reduction of eccentricity through various processes, 
{such as viscosity (e.g. \citealt{Ogilvie(2001),TeyssandierOgilvie(2016)}) or transport of angular momentum during the circularization process  (e.g. \citealt{Svirski(2017),BonnerotLu(2019)}), }
which are not self-consistently included in this work.
{Angular momentum may also be transported within the disk via secular processes, as the eccentric disk twists due to apsidal precession from internal pressure and general relativity, which requires more detailed additional calculations to understand (see eqs.~\ref{eq:dedt}-\ref{eq:dpomdt}).}
{We assume the disk eccentricity is set by the circularization of the most bound debris stream's eccentricity $\ef(\tfo)$, since the torque from the fallback {debris} falls off as $T_{\rm fb} \simeq -\dotMfb \sqrt{2 G \Mh \rp} \propto t^{-5/3}$.}
Figure~\ref{fig:ein} displays $\ein$ as a function of $\cV$, for different values of $\lam_e$.  We see $\ein \approx \lam_e$ when $\cV \lesssim 1$, while $\ein$ decreases to zero with increasing $\cV$ when $\cV \gtrsim 1$.  It may be shown that, when $\cV$ is given by equation~\eqref{eq:cVcirc}, the disk's inner eccentricity $e_{\rm in} = e_{\rm circ} = 0$, assuming that $\Mh \gg \Ms$ and $\bg$ is $\cO(1)$.

We need two additional boundary conditions to solve equation~\eqref{eq:e(a)}, which we take to be
\be
\left. \frac{\pd F}{\pd f} \right|_{a=\ain} = \left. \frac{\pd F}{\pd f} \right|_{a=\aout} = 0.
\label{eq:Ff_bdry}
\ee
We justify equation~\eqref{eq:Ff_bdry} in Appendix~\ref{sec:Ff_bdry}{, where we show this conidition is brought about by requiring equation~\eqref{eq:dpomdt_red} remain valid at the disk's boundaries when the radial pressure gradient is infinite.}  Our boundary conditions allow for an infinitesimally narrow eccentric disk to undergo apsidal precession: In the limit where $\aout \to \ain$, $\om$ does not vanish.  This effect has been reproduced analytically in other applications of narrow eccentric disks in the linear regime ($e \ll 1$; e.g. \citealt{ZanazziLai(2017)}).

\begin{figure}
	\includegraphics[width=\columnwidth]{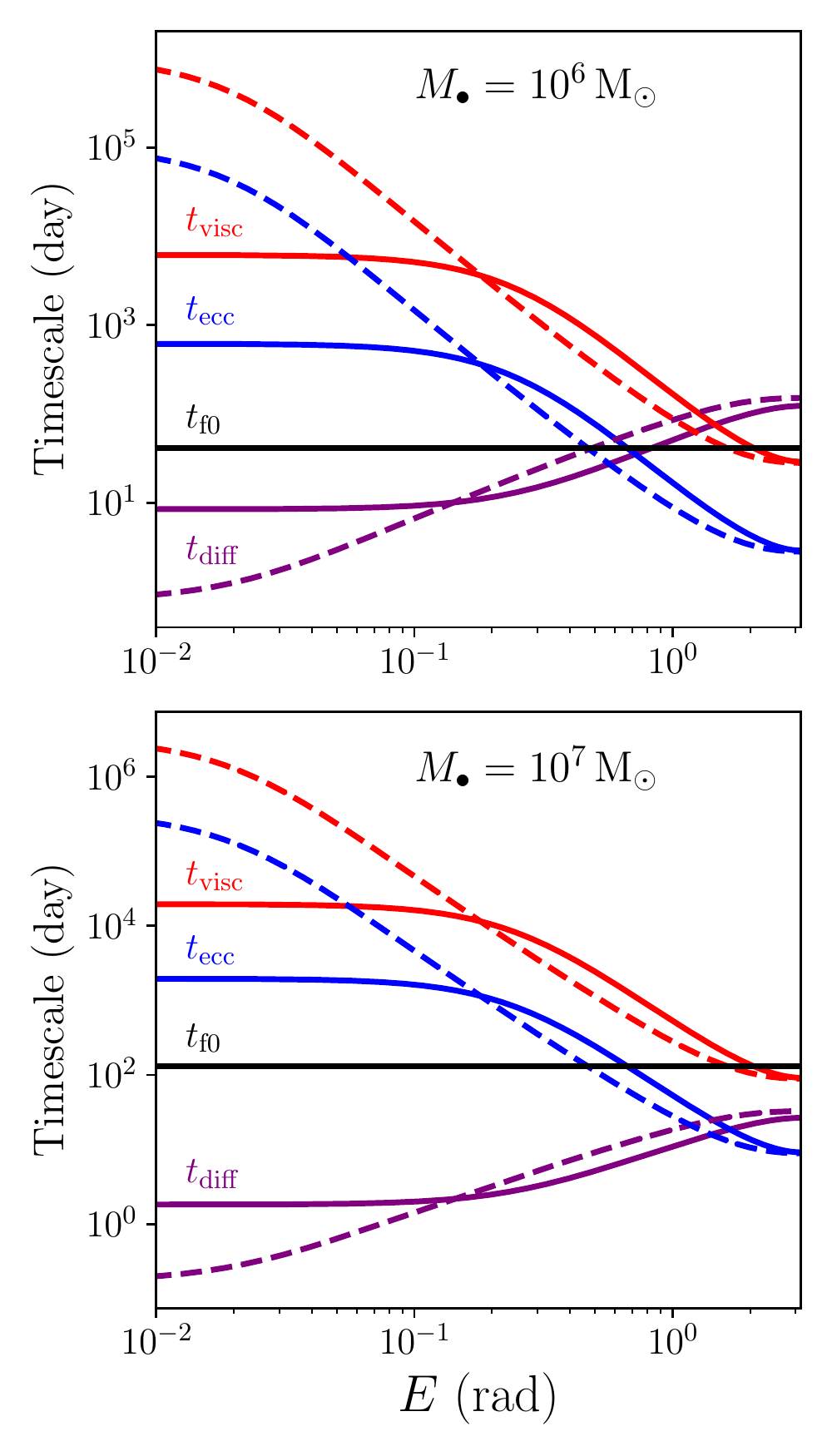}
    \caption{
    {
    {TDE lifetime $t_{\rm f0}$ (black, eq.~\ref{eq:tfo}), radiative diffusion $t_{\rm diff}$ (purple, eq.~\ref{eq:t_diff}), viscous $t_{\rm visc}$ (red, eq.~\ref{eq:t_visc}), and eccentricity propagation $t_{\rm ecc}$ (blue, eq.~\ref{eq:t_ecc}) timescales,} for $e = 0.5$ (solid lines) and $e = 0.7$ (dashed lines) and SMBH masses $\Mh$ indicated.  Here, $ae_a = a \pom_a = 0$, $\cV = 1$, $\Rs = \Rsun$, $\Ms = \Msun$, $a = \ain$, and $\ag = 0.1$.
    }
    }
    \label{fig:times}
\end{figure}

The formalism of \cite{OgilvieLynch(2019)} assumes an inviscid and adiabatic hydrodynamical disk, which is a valid approximation when the viscous time $t_{\rm visc} = a^2/\nu$ and vertical photon diffusion time $t_{\rm diff} = \tau H/c$ are much longer than the communication timescale for an eccentric disturbance $t_{\rm ecc} \sim (a/H)^2 n^{-1}$ \citep[e.g.][]{Ogilvie(2001),Ogilvie(2008),TeyssandierOgilvie(2016)}.  Assuming an $\ag$ viscosity ($\nu = \ag H^2 n$), these timescales work out to be
\begin{align}
t_{\rm diff} &= 34.5 \, \cV^{1/2} (1+\cV)^{1/2} \frac{\bMs^{5/3}}{\bRs \bMh^{2/3}} \frac{h}{j \ta^2} \ {\rm days},
\label{eq:t_diff} \\
t_{\rm visc} &= \frac{196}{\cV(1+\cV)^{1/2}} \left( \frac{0.1}{\ag} \right) \frac{\bRs^{3/2} \bMh^{1/2}}{\bMs} \frac{\ta^{3/2}}{h^2} \ {\rm days},
\label{eq:t_visc}
\end{align}
while
\be
t_{\rm ecc} \sim \frac{19.6}{\cV (1+\cV)^{1/2}} \frac{\bRs^{3/2} \bMh^{1/2}}{\bMs} \frac{\ta^{3/2}}{h^2} \ {\rm days}.
\label{eq:t_ecc}
\ee
{
{We note the timescale $t_{\rm ecc}$ was formally derived after averaging over the disk's azimuth (averaging over $E$), so it is not clear if it is the correct quantity to describe azimuthal variations of the eccentric disturbance propagation timescale.  We continue to use equation~\eqref{eq:t_ecc} with caution.}  In addition, TDE lifetimes scale with the orbital period of the most bound debris stream $\tfo$ (eq.~\ref{eq:tfo}).  Figure~\ref{fig:times} plots these timescales over a single orbit, varying $e$ and the SMBH mass $\Mh$.  Near apocenter ($E \gtrsim 0.3-1.0$), we see eccentric disturbances propagate at timescales shorter than typical TDE lifetimes ($t_{\rm ecc} \lesssim \tfo$), and approximating the disk as inviscid and adiabatic is reasonable ($t_{\rm ecc} \lesssim t_{\rm visc}, t_{\rm diff}$).  Our model of how the TDE disk propagates eccentric disturbances may break down near pericenter ($E \lesssim 0.3-1.0$), since the radial communication time near pericenter is longer than typical TDE lifetimes ($t_{\rm ecc} \gtrsim \tfo$).  Approximating the disk as inviscid will always be reasonable since $t_{\rm visc} \gtrsim t_{\rm ecc}, \tfo$ everywhere.  

In our work, we aim to calculate the peak optical and X-ray luminosities from thermal radiation before the disk has had time to cool ($t_{\rm diff} \gtrsim {\rm few} \ \tfo$).
For this reason, we assume the disk's internal energy stays approximately constant over the TDE's lifetime, and neglect cooling from radiative diffusion.  Examining Figure~\ref{fig:times}, because over much of the parameter space of interest, the diffusion time is much shorter than the TDE lifetime ($t_{\rm diff} \ll \tfo$), cooling from radiative diffusion is not negligible.  In this sense, our model is not self-consistent, and may cause us to overestimate the luminosity of TDEs.  Our approximations are worst near pericenter ($E \approx 0$) of highly eccentric disks, and for disks orbiting high-mass SMBHs ($\Mh \gtrsim 10^7 \, \Msun$).  
However, high-mass SMBHs are the most likely to circularize efficiently from stream-stream collisions near apocenter (see Fig.~\ref{fig:cV_plot}), so it is unclear if poorly circularized TDE disks can describe the optical emission of massive SMBHs in the first place.  We continue with our highly idealized TDE model, note our approximations are most valid for the thermal radiation of eccentric TDE disks near apocenter orbiting low-mass SMBHs, and caution the interpretation of our results beyond the order-of-magnitude level.  Specifically, our model likely overestimates the luminosity of high-energy thermal emission near pericenter (X-ray emission, see Sec.~\ref{sec:TDE_Emit}), as well as the bolometric luminosity of high-mass SMBHs ($\Mh \gtrsim 10^7 \, \Msun$).
}

\subsection{Eccentric TDE Disk Solutions}
\label{sec:EccTDEDiskSols}

\begin{figure}
	\includegraphics[width=\columnwidth]{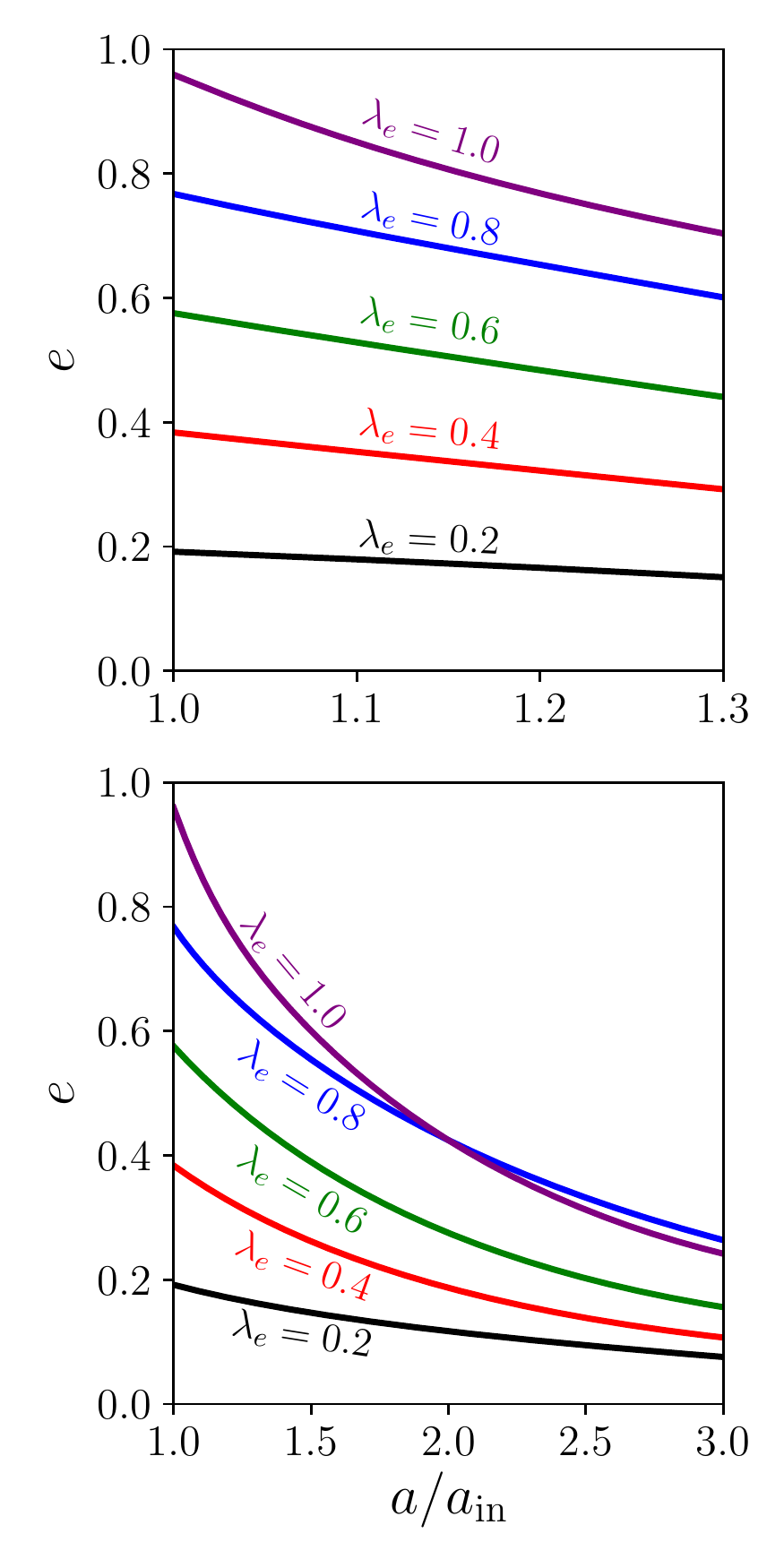}
    \caption{
    Disk eccentric modes $e$ as functions of semi-major axis $a$, with the outer semi-major axis values of $a_{\rm out} = 1.3 \, \ain$ (top panel) and $\aout = 3 \, \ain$ (bottom panel), for the $\lam_e$ values (see eq.~\ref{eq:ein}) as indicated.  Dimensionless eigenfrequencies $\tom$ (eq.~\ref{eq:tom}) are $\tom = -0.71$ ($\lam_e=0.2$), $\tom = 0.12$ ($\lam_e = 0.4$), $\tom = 0.31$ ($\lam_e = 0.6$), $\tom = 0.02$ ($\lam_e = 0.8$), and $\tom = 0.57$ ($\lam_e = 1.0$) for $\ain = 1.3 \, \aout$, and $\tom = 0.13$ ($\lam_e = 0.2$), $\tom = 0.51$ ($\lam_e = 0.4$), $\tom = 0.56$ ($\lam_e = 0.6$), $\tom = 0.30$ ($\lam_e = 0.8$), and $\tom = 0.56$ ($\lam_e = 1.0$) for $\ain = 3 \, \aout$.  Here, $\Mh = 10^6 \, \Msun$, $\Ms = 1 \, \Msun$, $\Rs = 1 \, \Rsun$, and $\cV = 1$.
    }
    \label{fig:Ecc_Sols}
\end{figure}

\begin{figure*}
	\includegraphics[scale=0.69]{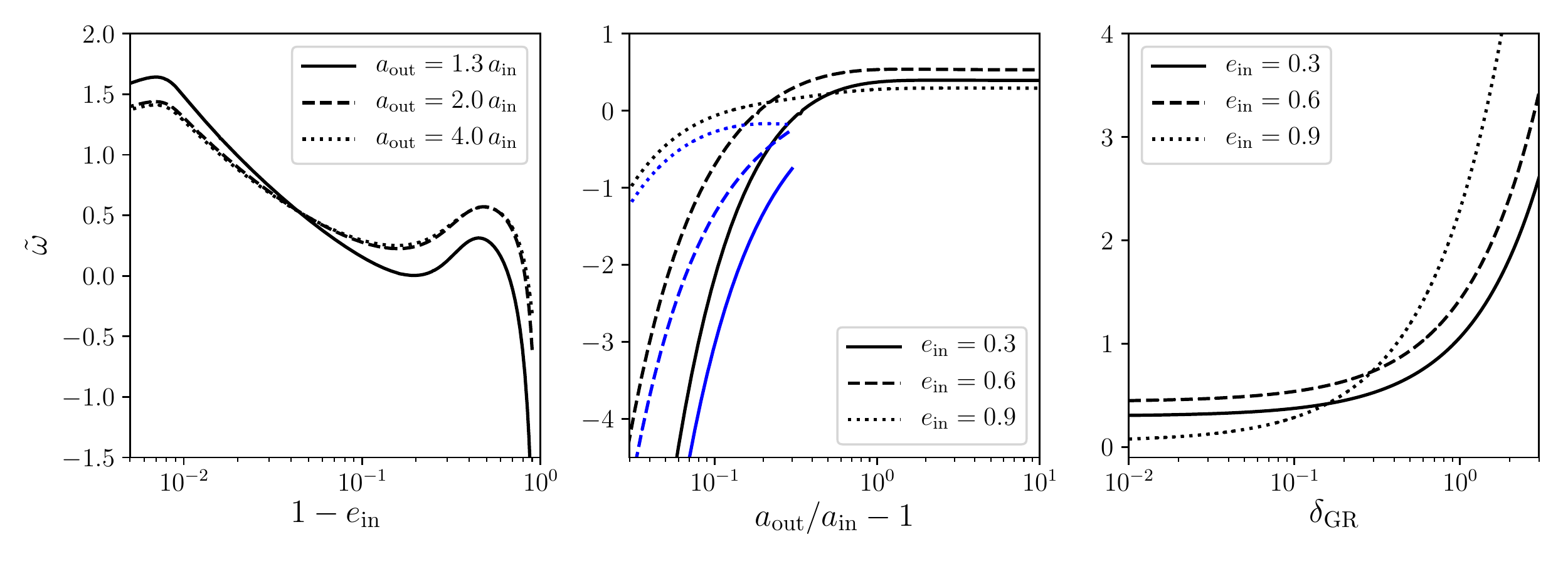}
    \caption{Dimensionless apsidal precession rate eigenfrequency $\tom$ (eq.~\ref{eq:tom}) as functions of the inner disk's eccentricity $e(\ain) = e_{\rm in}$ (left panel), annular extent of the disk $\aout/\ain-1$ (middle planel), and dimensionless apsidal precession rate from General Relativity $\dg_{\rm GR}$ (right panel; eq.~\ref{eq:dg_GR}), with solid, dashed, and dotted black lines denoting the outer disk semi-major axis $\aout$ (left panel) and $e_{\rm in}$ (middle \& right panels) values indicated.  Blue lines in the middle panel denote $\tom$ values calculated semi-analytically in the thin-annulus limit (App.~\ref{sec:Narrow_Disk}), with solid, dashed, and dotted lines denoting $e_{\rm in}$ values.  Since the only dependence on $\ain$ in equation~\eqref{eq:e(a)} is through $\dg_{\rm GR}$, varying $\dg_{\rm GR}$ is equivalent to varying $\ain$ with all other parameters fixed.  All plots take $\Mh = 10^6 \, \Msun$, $\Ms = 1 \, \Msun$, and $\Rs = 1 \, \Rsun$, with $\cV = 1$ ($\dg_{\rm GR} = 0.0306$; left \& middle panels) and $\aout = 2 \, \ain$ (right panel).}
    \label{fig:Ecc_Freqs}
\end{figure*}

This section investigates the lowest frequency, global solutions for our TDE disk model (Sec.~\ref{sec:TDEDiskStructure}).  We solve equation~\eqref{eq:e(a)} for the eccentric solution with no radial nodes {($e(a)>0$)}, calculating $e(a)$ and its corresponding eigenfrequency $\tom$ using the shooting method \citep{Press(2002)}.  We focus on the lowest-order, `fundamental mode' of the eccentric disk, because higher-order solutions (eccentricity profiles with more nodes, or more locations where $e(a) = 0$) are expected to have higher damping rates from dissipative processes, such as viscous damping \citep[e.g.][]{TeyssandierOgilvie(2016),Lee(2019a),Lee(2019b)}.  Our picture of how the disk evolves from the highly eccentric fallback debris orbits given by equation~\eqref{eq:ef} to the eccentricity profiles calculated in this section, is that a number of non-linear eccentric modes are excited soon after the TDE disk's formation, but the higher-order modes quickly damp their amplitudes, while the lowest-order apsidally-aligned fundamental mode remains excited over the TDE disk's lifetime.

Figure~\ref{fig:Ecc_Sols} shows a number of fundamental mode solutions using our disk model, for narrow ($\aout = 1.3 \, \ain$) and extended ($\aout = 3 \, \ain$) disks {(with $\aout/\ain$ values chosen arbitrarily to represent earlier and later phases of the disk's evolution)}, assuming different $\ein$ values (eq.~\ref{eq:ein}).  We see that highly eccentric, coherently precessing solutions are possible in TDE disks.  Although the details of each solution depend on the value of $\ein$, all solutions are qualitatively similar, and start off with large $e$ values near $\ain$ which decline monotonically as $a$ increases.  Our solutions differ qualitatively from the hydrodynamical simulations of \cite{Shiokawa(2015)}, who find $e$ increases as $a$ approaches $\aout$.  We believe this is due to the torque that fallback debris exerts on the outer disk which excites $e$ near $\aout$, whose effect we neglect in this work.
{It can be verified the orbital intersection parameter
\be
q^2 = \frac{(a e_a)^2 + (1-e^2)(a e \pom_a)^2}{[1 - e(e + a e_a)]^2}
\label{eq:q}
\ee
has a magnitude $|q|<1$ for the solutions displayed in Figure~\ref{fig:Ecc_Sols}, implying there are no orbital intersections \citep{OgilvieLynch(2019)}.
}

Figure~\ref{fig:Ecc_Freqs} displays the apsidal precession rate eigenfrequencies $\tom$ (global apsidal precession rate of eccentric disk) of our solutions, as functions of different parameters.  The left panel shows $\tom$ has a dependence on $\ein$, due to pressure inducing prograde rather than retrograde precession when the disk becomes highly eccentric (see \citealt{Ogilvie(2001),Ogilvie(2008),TeyssandierOgilvie(2016),OgilvieLynch(2019)} for discussion), independent of the disk's annular extent.  The middle panel shows $\tom$ increases as the disk becomes more radially extended for all $\ein$ values investigated, with $\tom$ asymptotically approaching a constant value when $\aout \gtrsim 2 \, \ain$.  Blue lines display $\tom$ calculated semi-analytically in the narrow disk-annulus limit ($\aout - \ain \ll \ain$, see App.~\ref{sec:Narrow_Disk}), which asymptotically approach the numerically calculated $\tom$ values using equation~\eqref{eq:e(a)}, indicating we have correctly calculated the lowest-order eccentric solution (smallest value of $|\tom|$).  
{The $\tom$ value diverges as $\aout \to \ain$ because, similar to p-modes in stars (see reviews by \citealt{Unno(1979),Christensen-Dalsgaard(2014)}), the eccentric eigenmode precession frequency is {inversely} proportional to the propagation time of a pressure wave $t_{\rm press} \sim (\aout-\ain)/\sqrt{\beps}$: as $\aout \to \ain$, $\tom \propto (\aout-\ain)^{-1} \to \infty$.}
Right panel displays the dependence of $\tom$ on the dimensionless apsidal precession rate from GR $\dg_{\rm GR}$ (eq.~\ref{eq:dg_GR}).  We see GR does little to modify $\tom$ when $\dg_{\rm GR} \lesssim 0.1$, but modifies $\tom$ when $\dg_{\rm GR} \gtrsim 0.1$, for the $\ein$ values investigated.  

Evaluating $\om$ in terms of $\tom$, we see
\be
\om = 4.07 \times 10^{-3} \, \left( \frac{\tom}{0.1} \right) \cV(1 + \cV)^{1/2} \frac{\bMs}{\bRs^{3/2} \bMh^{1/2}} \frac{2\pi}{{\rm days}}.
\label{eq:om_days}
\ee
This should be compared to the apsidal precession rate of the fallback debris stream due to GR, since a newly formed eccentric TDE disk will likely precess at a comparable rate:
\begin{align}
&\left. \frac{\pd \pom}{\pd t} \right|_{\rm f} = \frac{3 \Rg n(\af)}{\af (1 - \ef^2)} \simeq \frac{3 \Rg}{2 \rp} \frac{2\pi}{\tfo} \left( \frac{\tfo}{t} \right)
\nonumber \\
&= 7.77 \times 10^{-4} \, \bg \frac{\bMh^{1/6} \bMs^{4/3}}{\bRs^{5/2}} \left( \frac{\tfo}{t} \right) \frac{2\pi}{{\rm days}}.
\label{eq:dpomdt_f}
\end{align}
 Equation~\eqref{eq:dpomdt_f} is related to $\dg_{\rm GR}$ via $\pd \pom/\pd t|_{\rm f} = \dg_{\rm GR} [\ta (1-e^2)]^{-1} [\cH_a^\circ/(n a^2 M_a)]_{a=\ain}$, with $a = \af$ and $e = \ef$.  More detailed hydrodynamical simulations and analytical calculations of geodesic paths around the SMBH \citep[e.g.][]{Kochanek(1994),Hayasaki(2013),Hayasaki(2016),GuillochonRamirez-Ruiz(2013),GuillochonRamirez-Ruiz(2015),Dai(2013),Dai(2015),Bonnerot(2016),LuBonnerot(2020)} give comparable values of $\pd \pom/\pd t|_{\rm f}$.  Since $\om$ can be comparable to $\pd \pom/\pd t|_{\rm f}$ for the $\tom$ values we calculated in Figures~\ref{fig:Ecc_Sols} and~\ref{fig:Ecc_Freqs}, we argue it is feasible for a TDE disk to remain highly eccentric and coherently precess for a short amount of time after formation ($t \lesssim {\rm few} \ \tfo$).  The next sections investigate the thermal emission due to such highly eccentric disks.

\section{Eccentric TDE Disk Emission}
\label{sec:TDE_Emit}

This section calculates the emission from a highly eccentric TDE disk.  Section~\ref{sec:TDE_Emit_Est} gives order of magnitude estimates for the disk temperatures and luminosities expected from our theory, Section~\ref{sec:TDE_Emit_Model} introduces the model we use to calculate the eccentric disk's emission, while Section~\ref{sec:TDE_Emit_Result} presents our results for the emission from eccentric TDE disks.  

\subsection{TDE Emission Estimates}
\label{sec:TDE_Emit_Est}

In the `classical' model for TDE emission, an accretion disk around the SMBH rapidly forms after the TDE with an outer radius roughly equal to the circularization radius $r_{\rm out} \simeq 2 \rp$ \citep{Rees(1988),EvansKochanek(1989),Cannizzo(1990),Ulmer(1999)}.  Since the accretion rate from fallback debris $\der M_{\rm f}/\der t$ is expected to be super-Eddington (eq.~\ref{eq:dMfdt}), the TDE luminosity is expected to be of order the Eddington luminosity $L_{\rm Edd}$ at early times:
\be
L_{\rm Edd} = \frac{4\pi G \Mh c}{\kg} = 1.47 \times 10^{44} \, \bMh \, {\rm erg}\,{\rm s}^{-1}.
\label{eq:L_Edd}
\ee
The effective temperature of the TDE disk emission is then of order
\be
T_{\rm eff} \sim \left( \frac{L_{\rm Edd}}{4\pi \sg r_{\rm out}^2} \right)^{1/4} = 1.8 \times {10^5} \, \bg^{1/2} \frac{\bMh^{1/12} \bMs^{1/6}}{\bRs^{1/2}} \, {\rm K},
\label{eq:Teff_classic}
\ee
where $\sg$ is the {Stefan}-Boltzmann constant.  This simple model cannot explain the low effective temperatures ($\Teff \sim 2-3 \times {10^4} \, {\rm K}$) inferred from the optical emission of many TDEs \citep[e.g.][]{Komossa(2015),vanVelzen(2020)}.

In the shock-heating model of \cite{Shiokawa(2015),Piran(2015),Krolik(2016),Ryu(2020c)}, TDE emission is powered by shocks during the circularization process which heat the gas forming the accretion disk, and the disk never completely circularizes.  Instead of the TDE disk having a size of order the circularization radius, the size of the disk is of order the semi-major axis of the most tightly bound debris stream $a_{\rm f0} = G \Mh/(2 \Dg E) = \rp^2/(2 \Rs)$ (eq.~\ref{eq:ain}).  At late times ($t \gtrsim t_{\rm diff}$, see eq.~\ref{eq:t_diff}), the cooling from radiation can balance the shock-heating from the stellar debris, leading to a disk luminosity of order
\begin{align}
&L_{\rm shock} \sim \frac{\der \Mf}{\der t} \left( \frac{G \Mh}{a_{\rm f0}} \right)
\nonumber \\
&= 7.1 \times 10^{43} \, \frac{\bMs^{8/3}}{\bRs^{5/2} \bMh^{1/6}} \, {\rm erg}\,{\rm s}^{-1},
\label{eq:L_shock}
\end{align}
where we assume the velocity of the {bound} debris re-accreting onto the disk {is} of order $v_{\rm f} \sim (G \Mh/a_{\rm f0})^{1/2}$.  The effective temperature of the TDE disk emission is then of order
\be
T_{\rm eff} \sim \left( \frac{L_{\rm shock}}{4\pi \sg a_{\rm f0}^2} \right)^{1/4} = 3.0 \times {10^4}  \frac{\bMs}{\bRs^{9/8} \bMh^{3/8}} \left( \frac{\tfo}{t} \right)^{5/12} \, {\rm K},
\label{eq:Teff_shock}
\ee
a value much lower than the classical model for TDE emission (eq.~\ref{eq:Teff_classic}).

However, for our highly idealized model, we aim to calculate the TDE emission from the poorly circularized, eccentric disk at early times ($t \lesssim t_{\rm diff}$), before the disk has radiatively cooled.  Assuming energy transport occurs through radiative diffusion (see Sec.~\ref{sec:TDE_Emit_Model} for justification), the effective temperature for a poorly circularized disk is $T_{\rm eff}^\circ \simeq [4/(3\tau^\circ)]^{1/4} T_0^\circ$ (see eq.~\ref{eq:Teff} below), leading to a bolometric luminosity of order
\be
L \sim 4\pi \sg a_{\rm f0}^2 {T_{\rm eff}^\circ}^4 \sim 4.3 \times 10^{43} \left( \frac{\cV}{1+\cV} \right)^{1/2} \bMh \, {\rm erg}\,{\rm s}^{-1}.
\label{eq:L_model}
\ee
The following sections calculate the emission from our eccentric TDE disk models in more detail.

\subsection{Eccentric TDE Disk Emission Model}
\label{sec:TDE_Emit_Model}

\begin{figure}
	\includegraphics[width=\columnwidth]{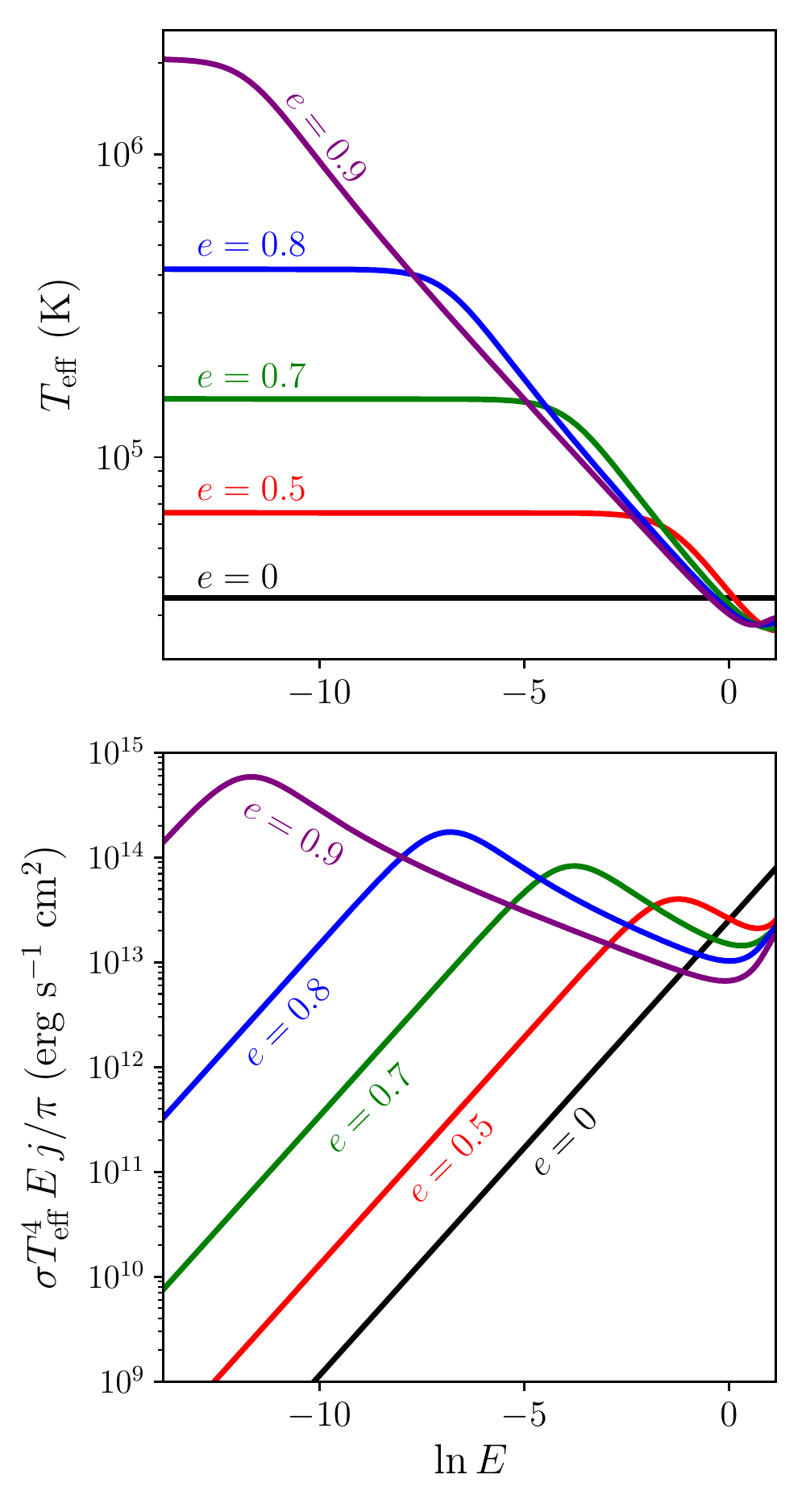}
    \caption{
    Effective temperature $\Teff$ (top panel; eq.~\ref{eq:Teff}) and flux per {logarithmic} unit eccentric anomaly $\ln E$,  $\sg \Teff^4 \,E \, j/\pi$ (bottom panel; see eq.~\ref{eq:L_bol}), as functions of $\ln E$, evaluated at the disk's inner semi-major axis $\ain$ (eq.~\ref{eq:ain}) for the eccentricity values $e$ indicated.  Here, $\Mh = 10^6 \, \Msun$, $\Ms = 1 \, \Msun$, $\Rs = 1 \, \Rsun$, $\cV=1$, and $a e_a = 0$.
    }
    \label{fig:Teff(E)}
\end{figure}

\begin{figure}
	\includegraphics[width=\columnwidth]{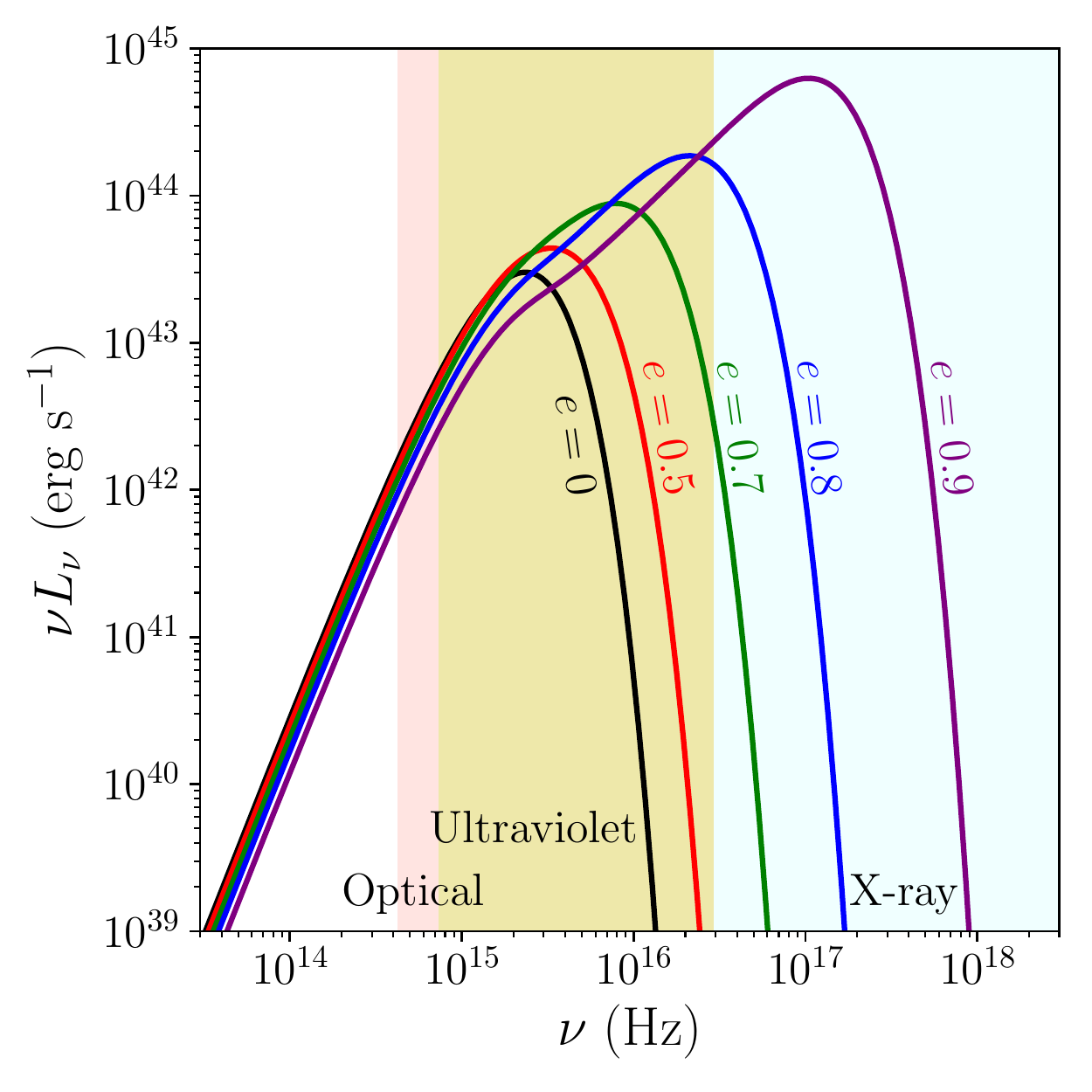}
    \caption{
    SEDs of emission from TDE disks with the constant eccentricity values $e$ indicated, with Optical, Ultraviolet, and X-ray frequency bands shaded as indicated.  Here, $\Mh = 10^6 \, \Msun$, $\Ms = 1\, \Msun$, $\Rs = 1 \, \Rsun$, $\cV = 1$, $a e_a = 0$, and $\aout = 2 \, \ain$.
    }
    \label{fig:Ecc_SED}
\end{figure}

When energy transport occurs through radiation, one can assume local thermodynamic equilibrium to calculate the effective temperature of the disk's emerging radiation \citep[e.g.][]{ZhuNarayan(2013)}:
\begin{align}
&T_{\rm eff} \simeq \left( \frac{4}{3 \tau} \right)^{1/4} T_0
\nonumber \\
&= 2.65 \times {10^4} \frac{\bMs^{1/3}}{\bMh^{1/12} \bRs^{1/2}} \frac{\cV^{1/8} (1 + \cV)^{3/8}}{\ta^{1/2} j^{1/12} h^{1/3}} \, {\rm K}.
\label{eq:Teff}
\end{align}
The top panel of Figure~\ref{fig:Teff(E)} displays $\Teff$ as a function of the {natural logarithm of the} disk's eccentric anomaly $\ln E$ at the disk's inner edge, for different eccentricity values $e$.  When $e$ is low, most of the disk emits with a temperature whose emission peaks in the optical or near UV ($\Teff \lesssim {10^5} \, {\rm K}$).  But when $e$ is high, the disk can emit with a temperature whose emission peaks in the far UV or X-ray ($\Teff \gtrsim {10^5} \, {\rm K}$) near $E \approx 0$ due to compressional heating near periapsis.

Equation~\eqref{eq:Teff} assumes energy transport is dominated by radiative diffusion, which requires the disk to be convectively stable: the thermodynamic gradient from radiation $\nabla_{\rm rd} = \der \ln T/\der \ln p|_{\rm rd} = -\left( \frac{4 \sg}{3 \rho \kg} \right) \der T^4/\der z$ must be smaller than the adiabatic thermodynamic gradient $\nabla_{\rm ad} = 1/4$.  Although one can calculate $\nabla_{\rm rd}$ for our disk model (see e.g. \citealt{ZhuNarayan(2013)}), this calculation is complicated by the fact that an eccentric disk does not lie in hydrostatic equilibrium ($\der p/\der z \ne -\rho n^2 z$).  Moreover, the highly turbulent formation of the eccentric TDE disk will significantly alter the disk's initial thermodynamic gradient.  Since hydrodynamical simulations are needed to accurately calculate the TDE disk's initial vertical structure, we continue to use the simple estimate~\eqref{eq:Teff} for the effective temperature.  Preliminary calculations carried out by the authors show a disk with energy transport dominated by convection ($\nabla \simeq \nabla_{\rm ad}$) increases $\Teff$ by a factor of $\sim 2$.  We note the hydrodynamical simulations of \cite{Sadowski(2016)} found convectively stable and unstable regions in a newly formed TDE disk.

{
In addition, equation~\eqref{eq:Teff} implicitly assumes Local Thermodynamic Equilibrium (LTE), which requires a diffusion time shorter than the lifetime of the TDE disk ($t_{\rm diff} \lesssim \tfo$).  This should be valid near pericenter for most TDEs, as well as TDEs around massive ($\Mh \gtrsim 10^7 \, \Msun$) SMBHs, but will break down near apocenter for low-mass SMBHs ($\Mh \sim 10^6 \, \Msun$, see Fig.~\ref{fig:times}).  However, because we are primarily interested in the accretion disk's photosphere blackbody temperature (which almost always transports energy primarily via radiative diffusion, see e.g. \citealt{Rafikov(2007),ZhuNarayan(2013)}), and the timescale for the photosphere to reach LTE is much shorter than that of the entire disk's vertical extent (since $\tau \gg 1$, see eq.~\ref{eq:tau}), we argue the diffusion approximation (eq.~\ref{eq:Teff}) is reasonable.
}

To calculate the Spectral Energy Distribution (SED) from the eccentric disk, we assume the disk emits locally like a black-body with temperature $\Teff$ given by equation~\eqref{eq:Teff}, giving the luminosity per unit frequency $L_\nu$ of
\begin{align}
\nu L_\nu = 4\pi \int \nu B_\nu(\Teff) \, \der A
= 4\pi \int_0^{2\pi} \int_{\ain}^{\aout} \nu B_\nu \, j \, \der a \, \der E,
\label{eq:SED}
\end{align}
where $B_\nu(T)$ is the Planck function, using the notation of \cite{RybickiLightman(1979)}.  The bolometric luminosity from the disk is then
\be
L = 4 \sg \int_0^{2\pi} \int_{\ain}^{\aout} \Teff^4 \, j \, \der a \, \der E.
\label{eq:L_bol}
\ee
The bottom panel of Figure~\ref{fig:Teff(E)} displays the emitted flux per unit $\ln E$ at the disk's inner edge, for different $e$ values.  Although the emission area decreases near periapsis in an eccentric disk ($j$ small near $E \approx 0$), the flux near periapsis can still be many orders of magnitude larger than than the flux near apoapsis due to compressional heating ($h$ small near $E \approx 0$).  This makes it possible for the frequency-integrated  luminosity to be larger at periapsis than at apoapsis when the disk is eccentric.

Figure~\ref{fig:Ecc_SED} calculates the SED from TDE disks with constant eccentricity values {($e(a) = {\rm constant}$)}.  When $e$ is low, the SED peaks in the optical or near UV due to the disk's low $\Teff$.  But when $e$ is high, the compressionally heated disk material near periapsis can dominate the disk emission, and cause the SED to peak in the far UV and soft X-ray.  Furthermore, high $e$ values can raise the bolometric luminosity by orders of magnitude.

\subsection{Eccentric TDE Disk Emission Results}
\label{sec:TDE_Emit_Result}

\begin{figure*}
	\includegraphics[width=\textwidth]{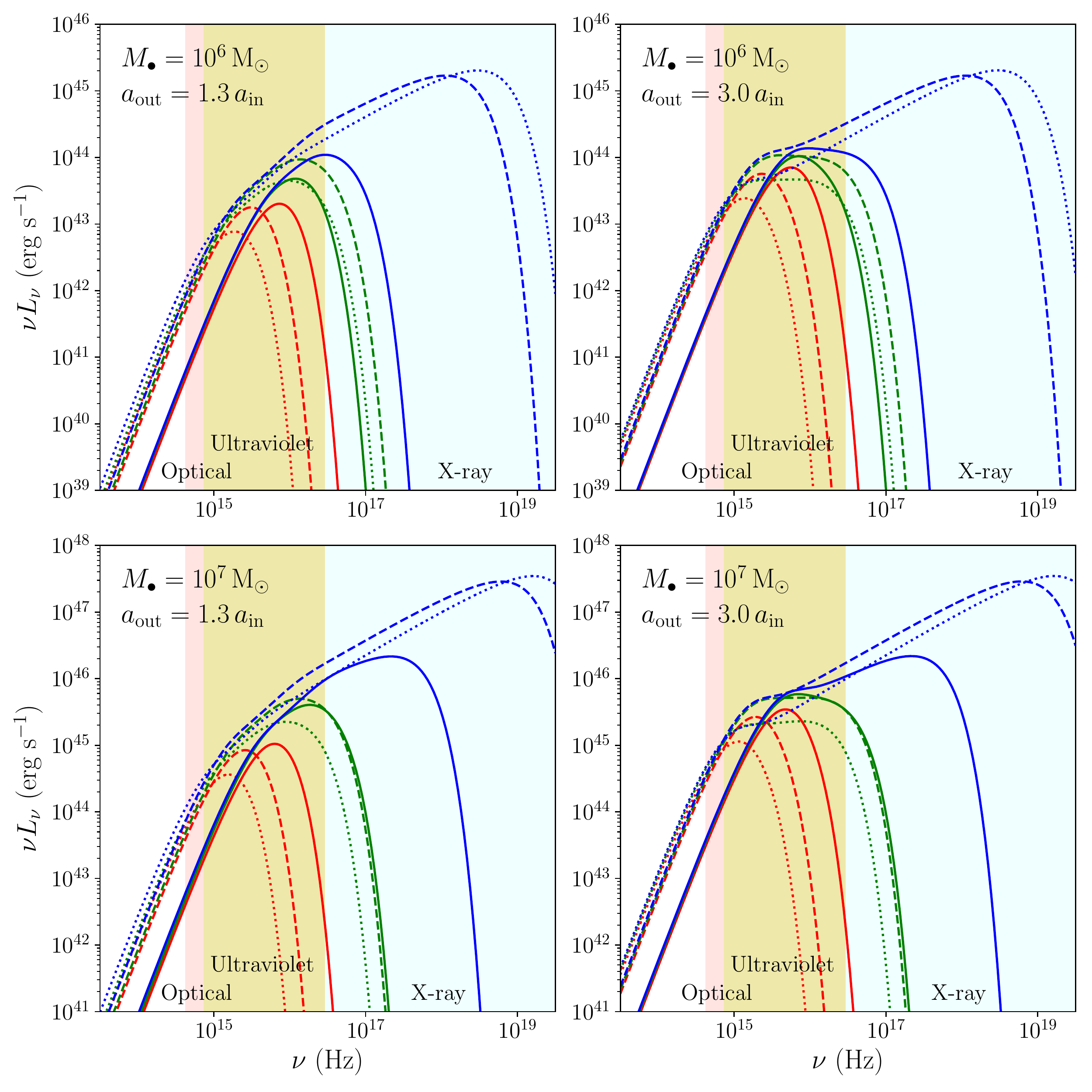}
    \caption{
    SEDs of emission from TDE disk eccentric eigenmodes (see Sec.~\ref{sec:EccTDEDiskSols}), with Optical, Ultraviolet, and X-ray frequency bands shaded as indicated.  Colors denote different $\lam_e$ values (eq.~\ref{eq:ein}), with $\lam_e = 0.4$ (red), $\lam_e = 0.8$ (green), and $\lam_e = 1.0$ (blue), while line styles denote different circularization efficiency parameter values $\cV$ (eqs.~\ref{eq:Eorb} \&~\ref{eq:epsInt}), with $\cV = 0.1$ (dotted), $\cV = 1$ (dashed), and $\cV = 10$ (solid).  Different panels display the SEDs for models with different SMBH masses $\Mh$ and outer disk semi-major axis $\aout$ as indicated.  Here, $\Ms = 1 \, \Msun$ and $\Rs = 1 \, \Rsun$.
    }
    \label{fig:Ecc_TDE_SEDs}
\end{figure*}

\begin{figure*}
	\includegraphics[width=\textwidth]{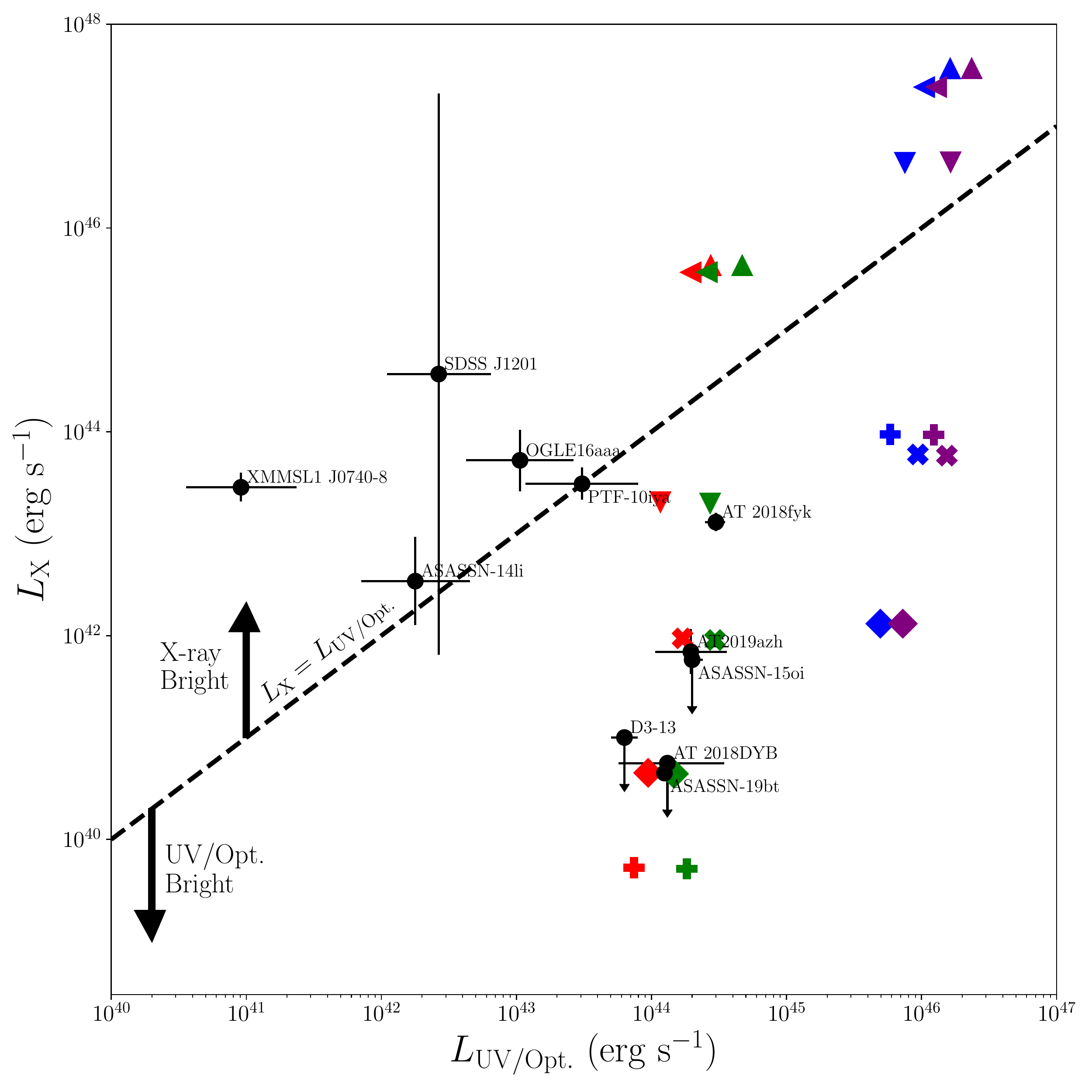}
    \caption{
    X-ray $L_{\rm X}$ (0.3-10 keV) verses Ultraviolet/Optical $L_{\rm UV/Opt.}$ (0.002-0.1 keV) luminosity for the eccentric eigenmode solutions from our disk models (colored symbols), compared to the estimated peak luminosity of observed non-jetted TDE candidates (black dots).  Symbol colors denote different normalized SMBH mass $\bMh$ (eq.~\ref{eq:norm}) and outer disk truncation semi-major axis $\taout = \aout/\ain$ values, with $(\bMh,\taout) = (1,1.3)$ (red), $(\bMh,\taout) = (1,3)$ (green), $(\bMh,\taout) = (10,1.3)$ (blue), and $(\bMh,\taout) = (10,3)$ (purple), while different symbols denote different inner disk eccentricity $\lam_e(\approx e_{\rm in})$ (eq.~\ref{eq:ein}) and circularization efficiency parameter $\cV$ (eqs.~\ref{eq:Eorb} \&~\ref{eq:epsInt}) values, with $(\lam_e,\cV) = (1,10)$ (triangle down), $(\lam_e,\cV) = (1,1)$ (triangle up), $(\lam_e,\cV) = (1,0.1)$ (triangle left), $(\lam_e,\cV) = (0.8,10)$ (plus), $(\lam_e,\cV) = (0.8,1)$ (x), $(\lam_e,\cV) = (0.8,0.1)$ (diamond).  Models with $\lam_e \lesssim 0.6$ have $L_{\rm X} < 3 \times 10^{38} \, {\rm erg}\,{\rm s}^{-1}$, and do not show on this plot.  Luminosity data from X-ray bright ($L_{\rm X} > L_{\rm UV/Opt.}$) TDEs SDSS J1201, XMMSL1 J0740-85, PTF-10iya, OGLE16aaa, ASASSN-14li, and NGC247 are from \protect\cite{Auchettl(2017)}, while UV/Opt. bright ($L_{\rm X} < L_{\rm UV/Opt.}$) TDE candidate data are from discovery papers: D3-13 \protect\citep{Gezari(2009)}, ASASSN-15oi \protect\citep{Holoien(2016)}, AT 2018DYB \protect\citep{Leloudas(2019)}, AT 2018fyk \protect\citep{Wevers(2019)}, ASASSN-19bt \protect\citep{Holoien(2019)}, and AT2019azh \protect\citep{Liu(2019),Hinkle(2020)}.  X-ray emission from D3-13, ASASSN-15oi, and AT 2018DYB was not detected: $L_{\rm X}$ values for these TDE candidates are upper limits on their peak X-ray luminosity values.
    }
    \label{fig:Obs_Comp}
\end{figure*}

This section uses the emission model of Section~\ref{sec:TDE_Emit_Model} to calculate the thermal emission from the eccentric TDE disk solutions from Section~\ref{sec:EccTDEDiskSols}.  Figure~\ref{fig:Ecc_TDE_SEDs} displays the SEDs for our eccentric disk solutions, varying parameters which set the inner disk eccentricity $\ein$ ($\lam_e$, see eq.~\ref{eq:ein}), efficiency of circularization $\cV$ (eqs.~\ref{eq:Eorb} \&~\ref{eq:epsInt}), annular extent $\aout-\ain$, and SMBH mass $\Mh$.  

The two parameters which have the largest effect on the frequency the disk SED peaks at (maximum $\nu L_\nu$ value) are $\lam_e$ (which affects $\ein$, see eq.~\ref{eq:ein}), and the efficiency of circularization parameter $\cV$ ({affecting $\ain$ and $\ein$,} see eqs.~\ref{eq:Eorb} \&~\ref{eq:epsInt}).  We see increasing $\lam_e$ increases the SED peak frequency (compare blue, green, \& red curves in Fig.~\ref{fig:Ecc_TDE_SEDs}).  Disk SEDs with low $\lam_e$ values peak in the Optical and near UV, while high $\lam_e$ values peak in the far UV and even X-ray, due to the high effective temperature $\Teff$ (eq.~\ref{eq:Teff}) near periapsis for eccentric disks (Fig.~\ref{fig:Teff(E)}).  The effect of changing $\cV$ depends on the disk's $\lam_e$ value (compare dotted, dashed, and solid lines in Fig.~\ref{fig:Ecc_TDE_SEDs}).  When $\lam_e$ is low ($\lam_e \lesssim 0.6$), increasing $\cV$ increases the SED peak frequency, since a larger $\cV$ makes the disk hotter (see eq.~\ref{eq:Teff}).  When $\lam_e$ is high ($\lam_e \gtrsim 0.6$), increasing $\cV$ can decrease the peak SED frequency.  This is because for large $\lam_e$, changing $\cV$ can have a significant impact on the disk's $\ein$ value (see Fig.~\ref{fig:ein}).

The shape of the TDE disk SED is primarily set by the eccentricity profile.  {In classical accretion disk models, the disk SED is found by summing over a number of {axisymmetric} rings, and the SED's shape is determined by the variation of $\Teff$ over the disk's annular extent \citep[e.g.][]{ChiangGoldreich(1997)}.  For our eccentric TDE disks, because our disks are narrow ($\aout/\ain \lesssim {\rm few}$) and the $\Teff$ dependence on $a$ is weak ($\Teff \propto a^{-1/2}$, see eq.~\ref{eq:Teff}), eccentric TDE disk SEDs are determined by how $\Teff$ varies over a single orbit from pericenter compression and apocenter expansion.  The SED shape is then determined by summing over a number of blackbodies at different azimuths $E$, which have very different temperatures.  In more detail,}   the low-frequency portion of the disk SED can be described by the Rayleigh-Jeans limit of the outer disk apoapsis emission ($\nu L_\nu \propto \nu^3$ when $h_{\rm P} \nu \ll k_{\rm B} T_{\rm eff}[\aout,\pi]$, where $h_{\rm P}$ is the Planck constant), while the high-frequency portion is the Wien limit of the inner disk's periapsis emission ($\nu L_\nu \propto \exp[-h_{\rm P}\nu/k_{\rm B} \Teff(\ain,0)]$).  Since $\Teff$ does not sensitively depend on semi-major axis $a$, the SED at mid frequencies ($k_{\rm B} \Teff[\aout,\pi] \lesssim h_{\rm P} \nu \lesssim k_{\rm B} \Teff[\ain,0]$) depends primarily on the eccentricity profile, since the magnitude and effective temperature of the flux near periapsis depends strongly on the disk eccentricity (Fig.~\ref{fig:Teff(E)}).  But although the shape of the SED is primarily set by the disk eccentricity profile, the annular extent of the disk has a small impact on the shape of the SED and total luminosity of the disk's low-energy (Optical \& near UV) emission.  Specifically, extended disks (high $\aout - \ain$) have more low-energy emission compared to narrow disks (low $\aout - \ain$; compare left \& right panels of Fig.~\ref{fig:Ecc_TDE_SEDs}), since $\Teff$ decreases with increasing $a$ in our model (see eq.~\ref{eq:Teff}).

The bolometric luminosity $L$ (eq.~\ref{eq:L_bol}) of the disk emission is primarily set by the SMBH mass $\Mh$, but also depends on the disk eccentricity through $\lam_e$.  Our simple scaling arguments (eq.~\ref{eq:L_model}) show $L$ from our TDE disk model should be comparable to and scale with $\Mh$ the same way as the Eddington luminosity (eq.~\ref{eq:L_Edd}). More detailed {calculations} of the disk SED find this is the case, with $L$ increasing by roughly an order of magnitude when $\Mh$ is increased from $10^6 \, \Msun$ to $10^7 \, \Msun$ (compare top and bottom panels in Fig.~\ref{fig:Ecc_TDE_SEDs}).  Increasing $\lam_e$ can also increase the X-ray luminosity by multiple orders of magnitude (compare red, green, \& blue curves in Fig.~\ref{fig:Ecc_TDE_SEDs}), due to the high $\Teff$ near periapsis in eccentric disks (Fig.~\ref{fig:Teff(E)}).  Scaling~\eqref{eq:L_model} also predicts $L \propto \cV^{1/2}$ when $\cV \lesssim 1$, which our SEDs roughly reproduce.

To compare the predictions of our TDE emission model with observations, Figure~\ref{fig:Obs_Comp} displays the peak X-ray $\Lx$ (0.3-10 keV) verses UV/Optical $\Luv$ (0.002-0.1 keV) luminosities of many TDE candidates (compare to Fig.~12 in \citealt{Auchettl(2017)}), and compares them to the $\Lx$ and $\Luv$ values from our models for different parameter values.  As a whole, we see our model tends to predict higher luminosities than the luminosities of both X-ray ($\Lx \gtrsim \Luv$) and Optically ($\Lx \lesssim \Luv$) bright TDEs, but does better at reproducing Optically bright TDE luminosities.  Like before, the main parameters controlling the properties of the TDE emission are the inner disk's eccentricity $\ein$ (through $\lam_e$ or $\cV$), the annular extent of the disk $\taout = \aout/\ain$, and the SMBH mass $\Mh$.  Increasing (decreasing) $\ein$ results in increasing (decreasing) $\Lx$ by orders of magnitude, with order unity (factor of $\sim 2-3$) corresponding changes in $\Luv$.  Increasing $\taout$ increases $\Luv$ with almost no change in $\Lx$, but an order of magnitude increase in $\taout - 1$ results in a factor of $\sim 2-3$ increase in $\Luv$.  Increasing $\Mh$ increases the bolometric luminosity $L \sim \Lx + \Luv$, as predicted by estimate~\eqref{eq:L_model}: Increasing $\Mh$ from $10^6 \, \Msun$ to $10^7 \, \Msun$ increases both $\Lx$ and $\Luv$ by a factor of $\sim 10-100$.  For our model to explain the observed TDE $\Lx$ and $\Luv$ values, we generally require low SMBH masses ($\Mh \sim 10^6 \, \Msun$) and moderate disk eccentricities ($\ein \lesssim 0.8$).

\section{Discussion}
\label{sec:Discuss}

\subsection{Theoretical Uncertainties}
\label{sec:ThryUncert}

In this work, we calculate the dynamics and thermal emission from a highly eccentric TDE disk, assuming the nested eccentric annuli have aligned periapsis directions (untwisted).  More detailed hydrodynamical simulations of newly formed TDE disks show that the disk material is highly eccentric, but also very twisted, with periapsis directions varying significantly over the disk's radial extent \citep[e.g.][]{Shiokawa(2015),Sadowski(2016),BonnerotLu(2019)}.  This twisting occurs when the disk forms with a precession rate and eccentricity profile different from the solutions calculated in Section~\ref{sec:EccTDEDiskSols}, and becomes twisted due to general-relativistic apsidal precession and pressure torques.  {Highly oscillatory eccentric disturbances, potentially including high twists,} may induce extra dissipation as travelling eccentric disturbances approach the disk's inner edge \citep{LynchOgilvie(2019)}.  A non-zero disk twist will cause the disk eccentricity magnitude $e$ to evolve in time (\citealt{OgilvieLynch(2019)}, see eq.~\ref{eq:dedt}), implying a realistic eccentric TDE disk's dynamics will differ substantially from the constant (in time) eccentricity profiles calculated in Section~\ref{sec:EccTDEDiskSols}.

Since our goal was to compare the eccentric disk's thermal emission to observed peak TDE luminosities, we have neglected the time evolution of the disk's internal energy.  This approximation should break down as the disk's internal energy is redistributed within the disk, and released via radiation or accretion.  Thermal diffusion will play an important role in the disk's temperature evolution early on (compare eqs.~\ref{eq:tfo} \&~\ref{eq:t_diff}), while viscosity/accretion will become relevant only much later (compare eqs.~\ref{eq:tfo} \&~\ref{eq:t_visc}) for highly eccentric TDE disks, although estimate~\eqref{eq:t_visc} neglects additional turbulence induced by parametric instabilities \citep{Papaloizou(2005b), Papaloizou(2005a),BarkerOgilvie(2014),WienkersOgilvie(2018)} and the Magnetorotational instability \citep{Chan(2018)} in eccentric disks.

Our SED calculations assume a face-on viewing geometry (eqs.~\ref{eq:tau} \&~\ref{eq:SED}), neglecting viewing angle effects.  Viewing an accretion disk's inner edge at a significant inclination can block high-energy radiation \citep[e.g.][]{Dai(2018),CurdNarayan(2019)}, which may significantly reduce the X-ray luminosities for our TDE disk model, especially since the disk aspect ratio is much smaller at periapsis than apoapsis (see Fig.~\ref{fig:h(E)}).  The eccentric disk models of \cite{Liu(2017),Cao(2018)} already invoke non-zero viewing inclinations to fit the H$\alpha$ and optical emission lines in the TDEs PTF09djl and ASSASN-14li.  Further studies of how TDE emission is affected by the observer's orientation with respect to the disk would be of interest.

\subsection{Observational Implications}
\label{sec:ObsImps}

Considering how idealized our eccentric TDE disk model is, it does well at reproducing the observed X-ray $\Lx$ and UV/Optical $\Luv$ luminosities for a number of TDE candidates (Fig.~\ref{fig:Obs_Comp}).  In general, our model slightly overestimates both $\Lx$ and $\Luv$, requiring low SMBH masses ($\Mh \sim 10^6 \, \Msun$) and moderate disk eccentricities ($\ein \lesssim 0.8$) to explain the observed Optical/UV-bright TDEs.  An interesting next step would be to calculate the thermal emission from newly formed eccentric TDE disks using hydrodynamical simulations \citep[e.g.][]{Shiokawa(2015),Sadowski(2016),BonnerotLu(2019)}, to see if the predicted luminosities match observed peak TDE luminosities (both $\Lx$ \&~$\Luv$) better.

In addition to matching observed peak $\Lx$ and $\Luv$ values, our TDE disk model also predicts bolometric luminosities $L$ of order the Eddington luminosity (eqs.~\ref{eq:L_Edd} \&~\ref{eq:L_model}), with $L$ proportional to the SMBH's mass (eq.~\ref{eq:L_model}, Figs.~\ref{fig:Ecc_TDE_SEDs}-\ref{fig:Obs_Comp}).  This correlation is consistent with a few studies, which show a tentative correlation between $L$ and the SMBH mass \citep{Wevers(2017),Wevers(2019b),vanVelzen(2020)}.

Our highly idealized eccentric disk solutions also predict the disk should precess with a period a bit longer than typical TDE fallback times (eqs.~\ref{eq:tfo} \&~\ref{eq:om_days}, Fig.~\ref{fig:Ecc_Freqs}).  Such precession may be observable in the disk's X-ray emission, since viewing angle effects may block or unveil the compressionally heated gas near periapsis (see Figs.~\ref{fig:h(E)} \&~\ref{fig:Teff(E)}).  However, hydrodynamical simulations are needed to understand if the disk coherently precesses, or becomes twisted over the TDE's lifetime (see Sec.~\ref{sec:ThryUncert} for discussion).

\section{Conclusions}
\label{sec:Conclude}

Motivated by TDE candidates with emission bright in the near UV and Optical, we investigate the dynamics and thermal emission of highly eccentric TDE disks, powered by the energy liberated during the circularization of the stellar debris on nearly parabolic orbits.  Section~\ref{sec:TDEDiskStructureDynamics} set up our model for a poorly circularized TDE disk, and studied the dynamics of highly eccentric disks.  By modelling the disk as a nested continuous sequence of elliptical orbits which communicate via pressure forces, we calculate special apsidally aligned and uniformly precessing eccentric disk solutions, where the twisting forces from pressure and General-Relativistic (GR) apsidal precession from the SMBH balance the inertial force from the disk's global precession frequency.  We find the disk solutions are significantly eccentric and non-oscillatory across the disk's annular extent (Fig.~\ref{fig:Ecc_Sols}), with global precession frequencies which can be comparable to the apsidal precession frequency of the elliptical debris stream from GR (Fig.~\ref{fig:Ecc_Freqs}, eqs.~\ref{eq:dpomdt_GR}, \ref{eq:tom} \&~\ref{eq:om_days}).

Section~\ref{sec:TDE_Emit} calculates the thermal emission from our eccentric TDE disk model, taking into account compressional heating (cooling) near periapsis (apoapsis; see Figs.~\ref{fig:Coos} \&~\ref{fig:Teff(E)}).  We find for our model, disks with low eccentricities ($\ein \lesssim 0.8$) have SEDs which peak in the near UV/Optical, but highly eccentric ($\ein \gtrsim 0.8$) disks can be X-ray bright due to the high-temperature disk emission near periapsis (Figs.~\ref{fig:Teff(E)}-\ref{fig:Obs_Comp}).  Comparing the X-ray and UV/Optical luminosities of our model to a number of TDE candidates, we find our model can generate emission consistent with the luminosities of many UV/Optically bright TDEs (Fig.~\ref{fig:Obs_Comp}).

This work takes a significant step to quantitatively calculate the thermal emission expected from the stream-stream collisions model for TDE emission \citep{Shiokawa(2015),Piran(2015),Krolik(2016)}.  Hydrodynamical simulations are needed to test the dynamics and thermal emission predicted from our highly idealized model of an eccentric TDE disk, to see if shock-heating of elliptical debris streams is the main way optically bright TDEs are powered.

\section*{Acknowledgements}

{We thank the anonymous referee, whose many comments significantly improved the quality and clarity of this work.}  JZ thanks Almog Yalinewich for introducing him to the stream-stream collisions model, and {Cl\'ement Bonnerot,} Eugene Chiang, Julian Krolik, Elliot Lynch, Nick Stone, and Yanqin Wu for useful discussions.  JZ was supported by a CITA Postdoctoral Fellowship and DAMTP David Crighton Fellowship.  GIO acknowledges the support of STFC through grant
ST/P000673/1.  The authors thank the Yukawa Institute for Theoretical Physics at Kyoto University. 
Discussions during the YITP workshop YITP-T-19-07 on International Molecule-type 
Workshop "Tidal Disruption Events: General Relativistic Transients" were useful to 
complete this work.




\bibliographystyle{mnras}
\bibliography{Ecc_TDE_Disk_v3} 




\appendix

\section{Justification of Eccentric Disk Boundary Condition}
\label{sec:Ff_bdry}

As discussed in Section~\ref{sec:EccDiskForm}, to solve equation~\eqref{eq:e(a)} for the eccentricity profile $e(a)$ and eigenfrequency $\tom$, a suitable boundary condition must be chosen.  This appendix justifies our choice of boundary condition~\eqref{eq:Ff_bdry}.  As in \cite{OgilvieLynch(2019)}, we begin by modifying $M_a$ and $\epsInt$ to include a taper function which declines rapidly to zero near the boundaries.  Specifically, we take
\be
M_a \to M_a T(a),
\hspace{5mm}
\epsInt \to \epsInt T(a),
\ee
where
\be
T(a) = \tanh \left( \frac{a - \ain}{w_{\rm in}} \right) \tanh \left( \frac{\aout - a}{w_{\rm out}} \right),
\ee
$w_{\rm in} = \dg_{\rm in} \ain$, $w_{\rm out} = \dg_{\rm out} \aout$, and it is assumed $\dg_{\rm in}, \dg_{\rm out} \ll 1$.  Rewriting equation~\eqref{eq:dpomdt_red} as (assuming $\pd \pom/\pd t|_{\rm ext} = 0$)
\begin{align}
&- \frac{\om M_a}{\cH_a^\circ T} \frac{n a^2 e}{\sqrt{1-e^2}} = \frac{\pd F}{\pd e} - a e_a \frac{\pd^2 F}{\pd e \pd f}
\nonumber \\
&- (2 a e_a - a^2 e_{aa}) \frac{\pd^2 F}{\pd f^2} - \left( \frac{\der \ln \cH^\circ_a}{\der \ln a} + 2 \frac{\der \ln T}{\der \ln a} \right) \frac{\pd F}{\pd f},
\end{align}
regularity at the boundaries requires
\be
\frac{\om n a^2}{\beps^\circ} \frac{e}{\sqrt{1-e^2}} \Bigg|_{a=\ain,\aout} = 2 a \frac{\der T}{\der a} \frac{\pd F}{\pd f} \Bigg|_{a=\ain,\aout},
\label{eq:e_regular}
\ee
as shown by \cite{OgilvieLynch(2019)}.  In the limit $\dg_{\rm in},\dg_{\rm out} \to 0$, $\der T/\der a \to \infty$ as $a \to \ain,\aout$, while the left hand side of equation~\eqref{eq:e_regular} remains finite.  Hence regularity requires equation~\eqref{eq:Ff_bdry} to be satisfied when $\dg_{\rm in}, \dg_{\rm out} \to 0$.

\section{Non-Linear Eccentric Solutions for a Disk with a Narrow Annular Extent}
\label{sec:Narrow_Disk}

To check the solutions calculated in Section~\ref{sec:EccTDEDiskSols} are the lowest-order eigenfunctions for the eccentricity profile of an untwisted disk $e(a)$, we also calculate the eigenfunctions in the narrow annulus limit ($\aout - \ain \ll \ain$), whose solution can be computed semi-analytically.  Specifically, we expand $e$ and $a e_a$ at $a = \ain$ to leading order in $\meps = (\aout-\ain)/\ain \ll 1$, and use the boundary conditions to calculate $\tom$.  The inner boundary conditions give $e(\ain) = \ein$ (eq.~\ref{eq:ein}), and $a e_a|_{a=\ain}$ through equation~\eqref{eq:Ff_bdry}.  We then use equation~\eqref{eq:e(a)} to calculate
\begin{align}
a^2 e_{aa} &\frac{\pd^2 F}{\pd f^2} \Bigg|_{a=\ain} = \left[ -2 a e_a \frac{\pd^2 F}{\pd f^2} +  \tom \ta^{3/2} \frac{e}{\sqrt{1-e^2}} + \frac{\pd F}{\pd e} \right.
\nonumber \\
&- \left. a e_a \frac{\pd^2 F}{\pd e \pd f} - \frac{\der \cH_a^\circ}{\der \ln a} \frac{\pd F}{\pd f} - \frac{\dg_{\rm GR}}{\ta} \frac{e}{(1-e^2)^{3/2}} \right]_{a = \ain}.
\label{eq:a2eaa(ain)}
\end{align}
To leading order in $\meps$, $\tom$ is then determined by requiring that
\begin{align}
    &e(\aout) = e(\ain) + \meps \,  a e_a \big|_{a=\ain}, \\
    &a e_a \big|_{a=\aout} = a e_a \big|_{a=\ain} + \meps \, a^2 e_{aa} \big|_{a=\ain},
\end{align}
satisfy equation~\eqref{eq:Ff_bdry}.


\bsp	
\label{lastpage}
\end{document}